\documentclass[manuscript, screen]{acmart}
\AtBeginDocument{%
  \providecommand\BibTeX{{%
    \normalfont B\kern-0.5em{\scshape i\kern-0.25em b}\kern-0.8em\TeX}}}

\setcopyright{acmcopyright}
\copyrightyear{2018}
\acmYear{2018}
\acmDOI{XXXXXXX.XXXXXXX}

\acmConference[Conference acronym 'XX]{Make sure to enter the correct
  conference title from your rights confirmation emai}{June 03--05,
  2018}{Woodstock, NY}
\acmBooktitle{Woodstock '18: ACM Symposium on Neural Gaze Detection,
 June 03--05, 2018, Woodstock, NY} 
\acmPrice{15.00}
\acmISBN{978-1-4503-XXXX-X/18/06}

\usepackage{booktabs}
\usepackage{subcaption}
\usepackage[normalem]{ulem}

\begin{document}

\title{LLM-based Smart Reply (LSR): Enhancing Collaborative Performance with ChatGPT-mediated Smart Reply System}

%
% The "author" command and its associated commands are used to define
% the authors and their affiliations.
% Of note is the shared affiliation of the first two authors, and the
% "authornote" and "authornotemark" commands
% used to denote shared contribution to the research.

\author{Ashish Bastola}
\authornote{Both authors contributed equally to this research.}
\email{abastol@clemson.edu}
\author{Hao Wang}
\authornotemark[1]
\email{hao9@clemson.edu}
\affiliation{%
  \institution{Clemson University}
  \streetaddress{P.O. Box 1212}
  \city{Clemson}
  \state{SC}
  \country{USA}
  \postcode{29631}
}

\author{Judsen Hembree}
\affiliation{%
    \institution{Clemson University}
    \city{Clemson}
    \state{SC}
    \country{USA}}
\email{jhembre@g.clemson.edu}

\author{Pooja Yadav}
\affiliation{%
    \institution{Clemson University}
    \city{Clemson}
    \state{SC}
    \country{USA}}
\email{poojay@g.clemson.edu}

\author{Zihao Gong}
\affiliation{% School of Cultural and Social Studies
    \institution{Tokai University}
    \city{Tokyo}
    \country{Japan}}
\email{0CPD1206@mail.u-tokai.ac.jp}

\author{Emma Dixon}
\affiliation{%
    \institution{Clemson University}
    \city{Clemson}
    \state{SC}
    \country{USA}}
\email{eschare@g.clemson.edu}

\author{Abolfazl Razi}
\affiliation{%
    \institution{Clemson University}
    \city{Clemson}
    \state{SC}
    \country{USA}}
\email{arazi@clemson.edu}

% \author{Christopher Flathmann}
% \affiliation{%
%     \institution{Clemson University}
%     \city{Clemson}
%     \state{SC}
%     \country{USA}}
% \email{cflathm@clemson.edu}

\author{Nathan McNeese}
\affiliation{%
    \institution{Clemson University}
    \city{Clemson}
    \state{SC}
    \country{USA}}
\email{mcneese@clemson.edu}

%
% By default, the full list of authors will be used in the page
% headers. Often, this list is too long, and will overlap
% other information printed in the page headers. This command allows
% the author to define a more concise list
% of authors' names for this purpose.


\begin{abstract}
Interactive user interfaces have increasingly explored AI's role in enhancing communication efficiency and productivity in collaborative tasks. The emergence of Large Language Models (LLMs) such as ChatGPT has revolutionized conversational agents, employing advanced deep learning techniques to generate context-aware, coherent, and personalized responses. Consequently, LLM-based AI assistants provide a more natural and efficient user experience across various scenarios. 
In this paper, we study how LLM models can be used to improve work efficiency in collaborative workplaces. Specifically, we present an LLM-based Smart Reply (LSR) system utilizing the ChatGPT to generate personalized responses in professional collaborative scenarios while adapting to context and communication style based on prior responses. Our two-step process involves generating a preliminary response type (e.g., Agree, Disagree) to provide a generalized direction for message generation, thus reducing response drafting time. We conducted an experiment where participants completed simulated work tasks involving a Dual N-back test and subtask scheduling through Google Calendar while interacting with co-workers. Our findings indicate that the proposed LSR reduces overall workload, as measured by the NASA TLX, and improves work performance and productivity in the N-back task. We also provide qualitative analysis based on participants' experiences, as well as design considerations to provide future directions for improving such implementations.
\end{abstract}

\maketitle

% This paper is submit to: CUI 2024
% https://cui.acm.org/2024/
% https://cui.acm.org/2024/submit/papers/

% Our format should refer to:
% https://arxiv.org/pdf/2301.10710.pdf
% https://dl.acm.org/doi/pdf/10.1145/3491102.3517554
% https://www.acm.org/publications/proceedings-template

% Questions:
% Do we need to make a hypothesis for the t test
% Significance in Performance

\section{Introduction}
%Effective communication is the cornerstone for productivity and collaboration in the fast-paced digital workplace. Among the myriad of tools to facilitate workplace communication, conversational agents (CAs) such as smart reply systems have emerged to be ubiquitous to many of today's digital platforms as a way to enhance the communication experience. 

Effective communication is crucial for productivity and collaboration in today's rapid digital workplace. Conversational agents like smart reply systems have become essential on many digital platforms, improving the communication experience. With recent developments in large language models (LLMs)\cite{achiam2023gpt, katz2024gpt, team2023gemini, touvron2023llama}, there have been various implementations in the development of user interfaces using their API endpoints\cite{wu2023autogen, cai2022context, goodman2022lampost, yuan2022wordcraft}. 
One still under-explored area of research includes the empirical study regarding smart-reply systems using these LLMs. Despite being widely adopted across major platforms, we observed these implementations often fail to prioritize user needs, thus significantly hindering their overall benefit to users.
%Even though most of the well-known platforms today implement them, in at least some form, with our observations and reviews, we presume most of these implementations lack prioritizing user needs which majorly impedes their contribution to users in general. 

% In our implementation, we identify key design considerations for these smart systems as they differ from traditional models. This includes assessing their integration into workplace communication and evaluating their impact on workload reduction, productivity, and performance.

%With our implementation, we identify some of these associated design considerations as these systems differ from their traditional counterparts. Thus, this incurs some major considerations for these systems to easily blend among the users, especially when considering workplace communication, while also validating whether these technologies will contribute to the user in terms of workload reduction, productivity, and performance.

For years, collaborative research has been focused on understanding and enhancing collaborative work practices by enhancing communication efficiency\cite{kamel1998applying}, productivity \cite{carasik1988case, convertino2007supporting} and enhancing knowledge through question answering\cite{yadav2019feedpal,srba2015askalot, enembreck2002personal} in various workplace scenarios. Recent works that incorporate AI-powered tools like chatbots and smart replies utilizing language models \cite{hancock2020ai} aim to optimize conversation management\cite{zhang2023impact, lewis2016appreciative, dwivedi2023so, teixeira2021interplay, kulkarni2019conversational}, streamline workflows\cite{cifci2023ai, li2022using, bidot2011using, bastola2024driving}, and promote smoother information exchange among collaborators. AI has proven to assist in various collaborative aspects, such as context-aware responses\cite{jiang2022beyond, santos2016toward}, decision-making support\cite{zarate2008collaborative, schaefer2019integrating, kunnathuvalappil2018artificial, bastola2023multi}, real-time language translation \cite{chen2022two, wilson2018collaborative, tien2017internet, epstein2015wanted}, as well as sentiment analysis \cite{kung2023performance, regan2024can} tasks and thus bear great potential in bring the huge share of contribution when it comes to enhancing user efforts. Integrating AI into collaborative work can enhance efficiency and productivity, but it's crucial to tackle privacy, reliability, and trust issues when designing AI-powered tools for these settings.

%While integrating AI in collaborative work can significantly improve conversation efficiency and productivity, it is also important to address challenges of privacy concerns, system reliability, and user trust when designing and implementing AI-powered tools in collaborative work environments.  
Earlier chatbots often failed to convincingly generate human-like responses. However, recent studies have shown that people, even with training, struggle to identify ChatGPT-generated text, achieving only around 50\% accuracy \cite{Ref:Humans_can't_tell}. In contrast to sophisticated models like ChatGPT, primitive smart replies offer limited, shorter responses \cite{Ref:Smart_Reply}. Despite these limitations, these low-level systems account for 10\% of all email responses, highlighting the demand for smart reply systems \cite{Ref:Smart_Reply}. 
In day-to-day work environments, numerous conversations involve routine matters such as scheduling or seeking permissions. 
Automating these exchanges using LLM-based AI technologies can provide realistic, human-like responses, effectively reducing the need for human involvement. 
% As AI improves, the generative text becomes more lifelike, allowing smart replies to be longer and more complex. 
% While these interactions may be brief, they demand a certain level of human engagement. 
% By using such technologies for handling mundane inquiries, users can conserve cognitive resources and allocate more time to their core work tasks, potentially leading to a substantial increase in productivity.

This work seeks to investigate the potential of LLM-based Smart Reply (LSR) to improve communication efficiency in daily work, focusing on enhancing responses in formal and informal conversations. 
By employing experiments using the combination of working simulations, smart reply system, and real-world scenario mapping as well as interviews and surveys, this study aims to evaluate the impact of integrating LSR on work performance, productivity, and other workload measures and comprehensively understand how the LLMs can enhance conversation in professional settings.
In our simulated environment, participants perform simulated work (N-back test) while interrupted by the co-workers' messages, then perform the event schedule on Google Calendar based on their conversation. The Google Calendar task involves rescheduling, adding, and deleting meetings in the participant's calendar. 

To reach these objectives, we conducted a mixed-methods evaluation of the LSR to answer the following research questions:

    \begin{itemize}
       
        \item  RQ1: Does the LSR impact work performance, productivity, and workload in a collaborative workplace?

This research question explores how LSR can reshape and optimize collaborative work dynamics. By examining the effects of the LSR on work-related outcomes, we gain a deeper understanding of its potential to enhance efficiency, boost communication, and alleviate cognitive workload in the workplace. Thus, we conduct comprehensive experiments and summarize three findings regarding work performance, productivity, and mental demand.

        \item RQ2: What are the key factors that impact the user experience during collaborative work?

This research question aims to understand the differences in user experience when LSR is introduced into the collaborative work dynamic. By conducting a comprehensive investigation that combines quantitative surveys and qualitative interviews, we aim to identify and analyze the critical factors that impact the user experience during collaborative work interactions. This knowledge is significant in informing the design and optimization of AI-driven tools to create more efficient and user-friendly collaborative workplaces. 
\end{itemize}

\section{Related Work}
Smart reply systems have evolved substantially, becoming integral in enhancing user experience in digital communication. Initially, Smart Compose\cite{Ref:Smart_compose,kannan2016smart} were based on basic blocks of LSTM and GRUs, making basic suggestions for email responses\cite{Ref:Smart_Reply}. However, the onset of more advanced Transformer-based models such as BERT\cite{devlin2018bert}, Gemini \cite{team2023gemini}, GPT-3.0 and GPT-4\cite{brown2020language, achiam2023gpt} revolutionized this domain, offering enhanced context understanding and coherent text generation. Consequently, these models have been widely integrated into various digital communication platforms, such as search engines and several messaging platforms. The landscape is further evolving with ChatGPT, moving us closer to realizing artificial general intelligence (AGI)\cite{bubeck2023sparks}, promising improvements in relevance and contextuality of automated replies. 
A critical analysis from the human interaction perspective, is needed to address challenges arising from the large-scale deployment of these intricate models specific to smart-reply systems, ensuring they cater effectively to the dynamic needs of workplace communication while safeguarding ethical and privacy norms.

    \subsection{User Interactions with Conventional Smart Reply Systems}
Based on our review, current research on smart replies reveals a notable gap, especially in the thorough assessment incorporating human side testing. 
A study by Henderson et al. \cite{henderson2017efficient} is instrumental in showcasing optimization for large-scale applications, yet it primarily accentuates mathematical constraints over user experience, failing to counter what users usually expect. Moosa et al.'s evaluation of Google’s Allo \cite{moosa2016smart} underscored the complexities involving emotions, relationships, and personalities in AI-generated messages, marking a shift towards a more human-centric assessment. A noteworthy study involved developing an AI chatbot as an information portal for the COVID-19 crisis. Deployed in a real-world setting, it connected experts and information seekers. The study evaluated the chatbot's effectiveness in aiding users to locate needed information\cite{xiao2023powering}. However, the trend in the literature gravitates towards task-specific optimizations and confined datasets, largely sidelining the intricate dynamics of user interaction. This has been a recurrent theme, echoing the need for evaluations that are both exhaustive and inclusive of the user experience dimension. Mieczkowski et al.'s study \cite{mieczkowski2021ai} delineates the potential ramifications of AI-mediated communication on interpersonal perception and addresses an interplay between technology and humans. Moreover, research shows that there exists a significant effect on communication relating to the shift of context as well as norms of communication \cite{hohenstein2023artificial, hancock2020ai} when embedding AI to ease conversations. 
Our implementation tries to assess these interactions and privacy-related concerns when employing these technologies on a large scale when specifically considering smart reply systems \cite{kim2023propile}, with the major difference being using a more intelligent language model. 

    \subsection{Conversational Agents in the Workplace}
Smart reply systems, as a major kind of conversational agent, can significantly enhance communication in the workplace by optimizing various aspects of interpersonal and organizational interactions\cite{wamba2020influence, shim2002past}. 
Aside from those benefits, studies have also explored the unintended influence of AI on human agency, revealing that AI affects content and behavior while highlighting the fluid and complex nature of human-machine agency transfer\cite{wenker2023wrote}. Even though AI, in general, has been regarded to improve productivity significantly in tasks of text completion such as documenting\cite{eloundou2023gpts, manyika2017future} and same for smart reply in workplace communication, there are certain ethical considerations that in some way restrict some of these benefits\cite{urquhart2022working}. 
Unlike previous smart reply systems, LLMs generate context-aware, coherent, and personalized responses that are much more powerful and can be tremendously helpful if proper regulations are brought into place\cite{kasneci2023chatgpt, srivastava2022beyond} to broaden the range of tasks. Keeping these in mind, we can imagine LSR's ability to deliver a deeper understanding of user input, adapt to different communication styles and tones, and handle complex and challenging conversations, which differs from general smart replies in terms of scale and learning approach. Research has shown that even traditional chatbots, with their capability to respond in a user-friendly manner, can significantly enhance a company's recruitment strategy\cite{nawaz2019artificial}. Conversations with these seemingly realistic bots have even demonstrated the ability to engage and attract customers so that they leave the website only after making a purchase\cite{krishnan2022impact}. We can only envision this improving exponentially, considering the powerful systems we currently possess. They've already been deployed in some form to support, mediate, and facilitate business communication\cite{hancock2020ai, getchell2022artificial} while also helping the users to be more productive and effective at their work\cite{pereira2023systematic}. 
Thus, our human-side evaluation of these systems aims to make them even more user-friendly while addressing their concerns in-depth.

\section{Method}

%In this study, participants engaged in workplace simulation scenarios conducted entirely online, eliminating the need for in-person attendance. The experimental setup involved interactions with co-workers characterized by distinct personas: Jeff (neutral), Tony (formal/strict), and Janine (casual/talkative). 

We conducted the study entirely online, eliminating the need for in-person attendance. The experimental setup involved interactions with co-workers characterized by distinct personas: Jeff (neutral), Tony (formal/strict), and Janine (casual/talkative).
% As the study being long in time, which may cause fatigue,

    \begin{figure}[h]
    \centering
    \includegraphics[width=0.8\columnwidth]{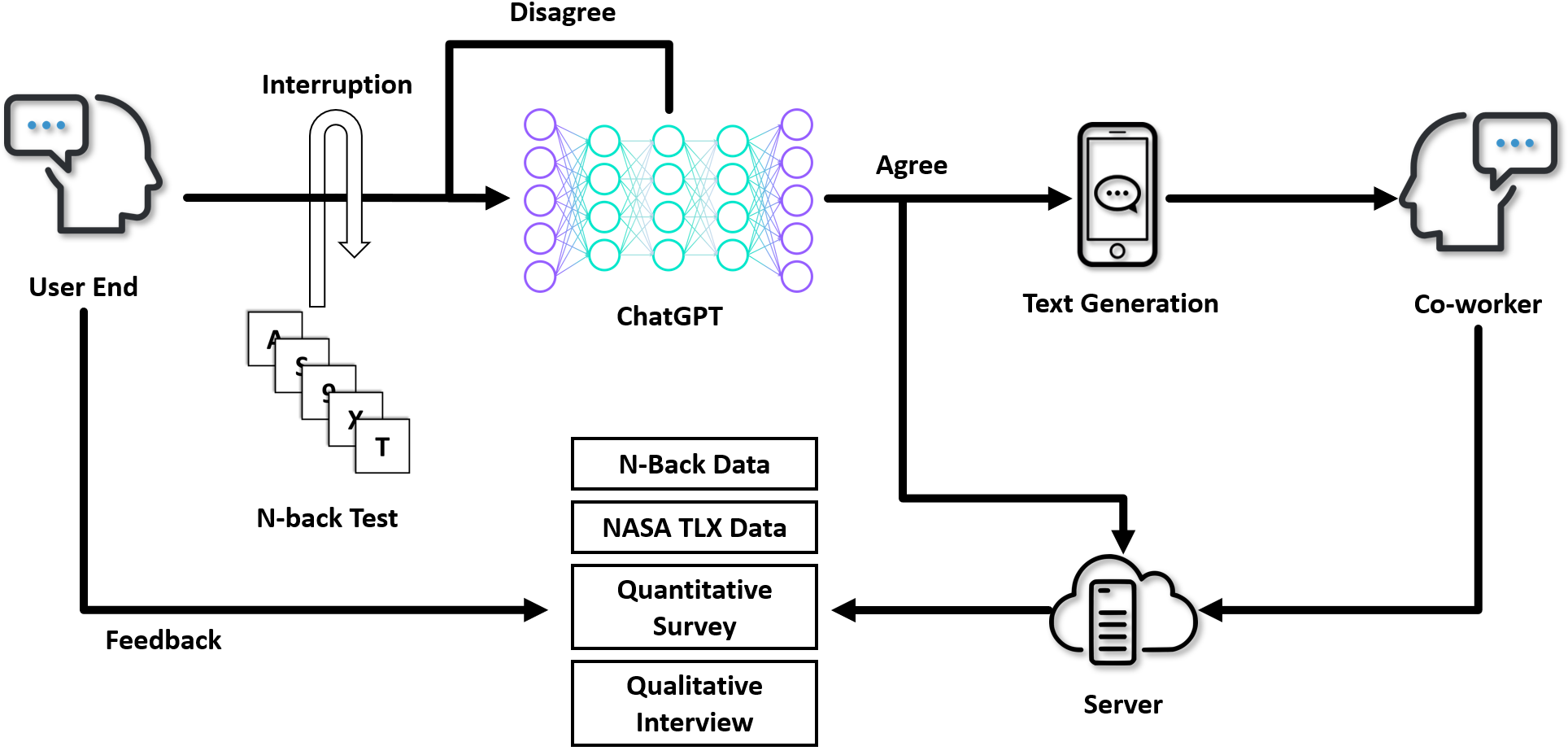}
    \caption{The overall workflow of experiment}
    \label{fig:workflow}
    \end{figure}

The actual experiment began by providing participants informed consent through Qualtrics. They then downloaded a designated program, watched an instructional video, and practiced the N-back test and Google Calendar scheduling until adept. After training, participants could seek clarification from the research team. Once ready, the official study began. For reference on the study's architecture, text generation, and data collection processes, please refer to Figure \ref{fig:workflow}.

In the initial phase, participants interacted with co-workers without LSR. Jeff then requested rescheduling meetings to different days, Janine introduced various queries and some informal conversations, and Tony acting as a boss, acted more formally(asking for progress updates). Participants then completed a NASA TLX questionnaire and had a short interview session to discuss their interactions. 
In the subsequent phase, participants repeated interactions with LSR's assistance. After participants completed the NASA TLX questionnaire and shared their thoughts in a voice chat session, they were asked to complete an exit survey and participate in a detailed interview about their experiences with LSR.

To ensure the fairness and accuracy of experiments, our experiments applied counterbalances on
(a) conversation timing (e.g., co-workers sending messages at random times), 
(b) conversation order (e.g., co-workers sending messages in a different order), 
and (c) conversation contents (e.g., vary the message expression without changing the purpose and the tone) 
to eliminate the learning effect of participants in the first part, prevent their attempt to predict the interruption of co-workers and the timing of assignment (e.g., schedule a meeting) in the second condition. 
Furthermore, participants were required to participate in a training session of around 15 to 20 minutes before the start of the experiment to make them fluent in N-back tasks, scheduling using Google Calendar, and typing messages manually to avoid carryover effects, if any.
% Considering our participants' extensive messaging experience, we deduced their typing and processing skills were already at saturation. 

\subsection{System Design}

\begin{figure}[htbp]
    \centering
    \includegraphics[width=0.8\columnwidth]{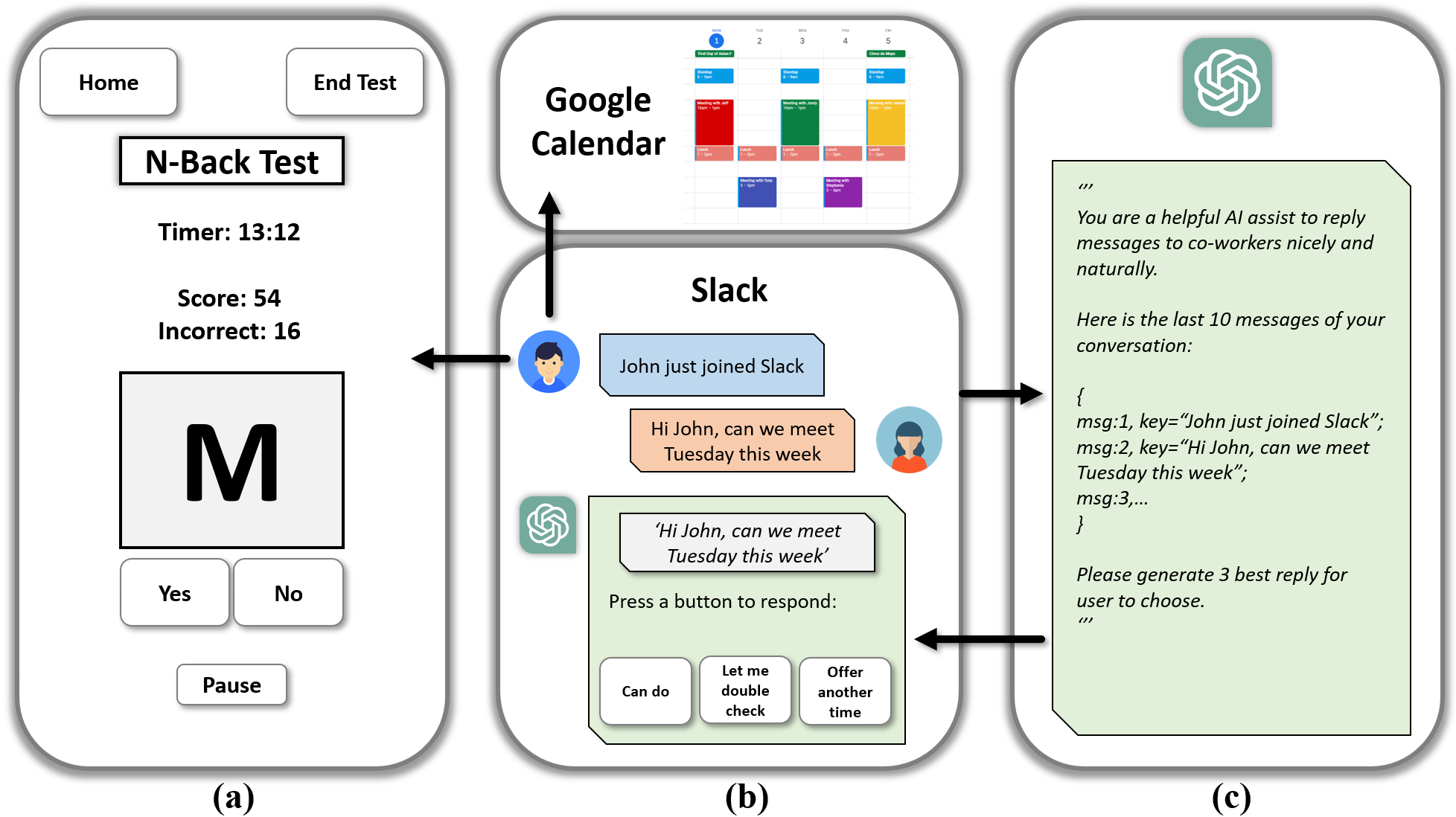}
    \caption{The main component of the proposed user interface. (a) N-Back test program; (b) Slack with LSR (top) and Google Calendar (bottom); (c) ChatGPT prompting template.}
    \label{fig:framework}
    \end{figure}

The implemented system consists of three main components: 1) the N-back test for simulated work, 2) subtask scheduling via Google Calendar, and 3) Slack integrated with LSR. The detailed design for the proposed system is depicted in Figure \ref{fig:framework}. 
% This system design aims to explore AI's impact on participant interactions in a simulated work environment, emphasizing the challenges and advantages of AI integration in such contexts.

\subsubsection{Simulated Work via the N-back Test}
To replicate real-world digital workplace challenges, we utilized the Dual N-Back test, a cognitively demanding task relying on working memory \cite{owen2005n}. 
% This task involves presenting participants with a sequence of stimuli (e.g. letters, numbers, or images) and instructing them to identify when the current item matches the item presented N steps back in the sequence. This choice aimed to simulate scenarios where diverting attention from a task to respond to a team member's inquiry could disrupt workflow and induce mental demand.

In our implementation, we specifically utilized a \textbf{2}-back test with alphabets, where participants were prompted to press a button whenever the current alphabet matched the alphabet presented \textbf{two} steps earlier. 
The N-back test was limited to \textbf{300} total letters, and data on total correct responses, total incorrect responses, and total time cost were recorded for each run in the Python back end. The implemented N-back program is a web-based application that was developed using HTML, Javascript, and Python. This offered adaptability across different platforms (desktops, tablets, and mobile phones) and could be customized for future task requirements. The user interface is depicted in Figure \ref{fig:framework}.

The proposed N-back program is open-source and accessible on GitHub:

% https://github.com/AIS-Clemson/Nback.   

        % \begin{figure}[htbp]
        % \centering
        % \includegraphics[width=0.6\columnwidth]{newImages/n-back.png}
        % \caption{A Screenshot of basic N-back implementation}
        % \label{fig:N_back}
        % \end{figure}

\subsubsection{Subtask Management Using Google Calendar}
During the N-back test, participants received work-related messages from simulated managers at unspecified, random intervals.
These messages replicated typical workday scenarios, including meeting rescheduling and casual exchanges, often necessitating schedule adjustments, which simulate the sudden events in the real world. 
Participants were required to temporarily pause their ongoing task (N-back test), to manage their schedules using a provided Google Calendar for necessary adjustments, as shown in Figure \ref{fig:framework}. Then, they need to respond to their co-workers' scheduling inquiries either manually (without LSR) or automatically (with LSR).

% This setup allowed us to assess how participants integrated work-related conversations and sudden events into their current work, providing insights into the impact of these interactions on their cognitive workload and efficiency.

\subsubsection{ChatGPT Integration With Slack}
The LSR integrates with Slack through its API. When a message is directed to a participant, it appears exclusively for them. Within this message, three response options are generated by sending the last ten messages of the ongoing conversation as a script to the ChatGPT. This approach allows for dynamic response generation. It's important to note that creating buttons for user selection is primarily AI-driven, with limited manual oversight or control. A new request is initiated once a user selects a button, and the AI generates a response based on the chosen response type. After selecting a button, the AI-generated message will be directly forwarded to the chat window on Slack. To enable this integration, we developed a server in Golang and provided remote access via Ngrok for seamless interaction with the Slack API. 
% It is noteworthy that our implementation intentionally provided a simple set of features to allow participants to explore and adapt based on their experiences rather than overwhelming them with options from the outset. This includes the exclusion of message editing functionality before sending. 

% This approach encouraged participants to identify this limitation and find alternative solutions, providing us with potential implementations that align with each participant's preferences. Features that garnered consensus among participants could inform better and more generalized interaction design implementations.

The proposed LSR system is open-source and accessible on GitHub:
% https://github.com/AIS-Clemson/LSR.
% add code later

\subsection{Participants}
This software development was initiated on January 25th, 2023. This study was approved by the Institutional Review Board (IRB) on April 18th, 2023. The executed comprehensive test workflow engaged a total of eighteen participants, including three pilot tests and fifteen full studies. 

    \begin{table}[h]
        \caption{Participant Demographics}
        \resizebox{1\columnwidth}{!}{%
        \begin{tabular}{cccccccc}
           \toprule
             Participant ID&Age& Gender & Job& Years of work&Slack experience  &Study Type  &English Speaker\\
                    \midrule
         T01& 30& Male& RA& 5& Never&Pilot
          &Y\\
         T02& 30& Male& GRA& 6& Weekly&Pilot
          &Y\\
         T03& 23& Male& Graduate Trainee& 1& Never&Pilot  &Y\\
             
             P01&26& Female& GRA& 2&Daily&Full  &Y\\
             P02&19& Female& Cashier& 3&Never&Full  &Y\\
             P03&57& Male& Paint Specialist& 40&Never &Full  &Y\\
             P04&26& Male& GRA& 4&Never &Full  &Y\\
             P05&27& Male& GRA& 4&Never&Full  &Y\\
             P06& 26& Male& TA& 2& Never &Full  &Y\\
             P07& 30& Male& GRA& 3& Monthly&Full  &Y\\
             P08& 25& Female& Product Engineer& 2& Never&Full  &Y\\
             P09& 34& Male& Teacher& 5& Never&Full  &Y\\
             P10& 25& Female& TA& 1& Never&Full  &Y\\
             P11& 23& Male& Graduate Engineering Trainee& 1& Never&Full  &Y\\
             P12& 26& Female& Software Engineer& 3& Rarely&Full  &Y\\
             P13& 30& Male& RA& 10& Never &Full  &Y\\
             P14& 30& Male& RA& 10& Never &Full  &Y\\
             P15& 24& Male& Mechanical Engineer& 1& Rarely &Full  &Y\\
             \bottomrule
        \end{tabular}
    }
    \label{TABLE:participants}
    \end{table}

Our participant pool was spanning individuals aged 19 to 57, with representation from both genders and various occupational roles. This diversity aims to eliminate potential biases linked to gender, age, and proficiency with AI technologies.

Participant demographic information is summarized in Table \ref{TABLE:participants}. All participants are represented by the testing ID assigned by the research team. Furthermore, all participants are recruited through the university email network, and all participants are English speakers, ensuring the minimum requirement of the test is fit.

% To ensure ethical and responsible research conduct, we obtained informed consent from each participant before the initiation of testing and interview. This step was necessary to address privacy concerns and uphold the highest standards of participant confidentiality and ethical research practices.

\subsection{Data Collection}

\subsubsection{Work Performance Tracking}
We evaluate work performance by assessing the correctness of tasks completed during the N-back test, both with LSR assistance and without, to measure the impact of LSR on the accuracy of work.

\subsubsection{Productivity Tracking}
Productivity was analyzed by calculating the messages per minute metric. This metric was determined by dividing the number of messages sent by a participant on Slack by the total time cost of the N-back test. This approach eliminates potential biases related to participants' proficiency in writing and sending messages on Slack and their proficiency in the N-back test. In essence, messages per minute provided a standardized measure of work efficiency, reflecting how promptly tasks were completed relative to resource cost (time).

\subsubsection{Cognitive Effort Tracking}
After each N-back test, we employ the NASA Task Load Index (NASA TLX) \cite{hart2006nasa}, a widely recognized method for assessing cognitive load. NASA TLX has become an essential instrument for understanding and enhancing the user experience among diverse domains, including human factors research, usability testing, and the evaluation of user interfaces. The main aspects of TLX are \textbf{1)} Mental Demand, \textbf{2)} Physical Demand, \textbf{3)} Temporal Demand, \textbf{4)} Performance, \textbf{5)} Efforts in the task, \textbf{6)} Frequency of similar situations, \textbf{7)} Frequency of similar situations, \textbf{8)} Affect on the temperament of other employees, and \textbf{9)} Frustration.

The NASA TLX evaluation in this study requires participants to provide ratings for each of its dimensions on a scale ranging from \textbf{0} (\textbf{very low} workload) to \textbf{20} (\textbf{very high} workload).

The individual scores derived from each dimension can be employed to generate a comprehensive workload score. 
This approach allowed us to assess the task's combined cognitive and physical demands and provided insights into participants' perceived cognitive workload and mental state.

     % Researchers and practitioners utilize this tool to ensure that systems, interfaces, and tasks are designed and assessed within acceptable workload thresholds, thus minimizing the potential for excessive mental or physical strain on users.

\subsubsection{User Experience Tracking}
After completing the two experimental rounds, participants were invited to complete a final survey comprising 10 questions focused on their experience with the LSR. The survey aimed to explore participants' expectations, emotions, and perceptions regarding the AI within the LSR system and their overall user experience with the LSR \cite{Ref:An_ideal_human}. Specifically, \textbf{Questions 1-6} assessed participants' preferences for AI using a five-point Likert scale, ranging from "\textbf{Strongly Disagree}" to "\textbf{Strongly Agree}". \textbf{Questions 7-10} then, seek user feedback on the implemented LSR, utilizing a five-point frequency scale from "\textbf{never}" to "\textbf{always}".

\subsubsection{Semi-structured Interviews}
In our study, interviews were conducted at three different times: after the first conversation with colleagues (without LSR), after the second conversation with colleagues  (with LSR), and after the participant finished both the NASA TLX and final surveys.  The interviews were open-ended and consisted of questions that aimed to understand the perceptions and feelings of participants towards the different co-workers and their ability to respond to their co-workers with and without the LSR system. Some example questions included in the interviews were: 

\begin{itemize}
    \item \textbf{How was your experience using LSR in your conversations?  
    \item Did the presence of LSR enhance your overall performance and productivity?
    \item How did you adapt to having an LSR? Did it require any adjustments to your usual way of working?  
    \item ...}

\end{itemize}

The semi-structured interviews allowed researchers to dive deeper into specific aspects of the conversations they deemed significant, encouraging participants to provide more detailed insights when their initial responses were brief. While we have provided a selection of questions here for context, it is important to note that the interviews and surveys formed a comprehensive data collection approach to connect participant experiences with our research questions. 

All interviews were audio-recorded and spot-transcribed by the research team to collect the most salient quotes from participants as they related to the research questions. 
To analyze the interview data, a deductive thematic analysis approach was taken \cite{braun2006using}. 
The first and second authors reviewed the interview data and coded the data based on the major categories of interest in this study: work performance, productivity, cognitive demand, impacts on team dynamics, opportunities for improvement of the AI teammate, and willingness to use AI teammate. These codes were then grouped into major themes, which the research team reviewed and iterated on. The themes were more clearly defined and refined, resulting in the three larger themes: "Effect of AI-assistant on Participants' Work Performance, Productivity, and Cognitive Demand", "Key factors that impacted the user's experience with the LSR", and "Opportunities for improvement of the LSR", each of which is detailed in Section \ref{result}.

\section{Results}
\label{result}
In this section, we report the results of our study as they relate to the effect of the LSR on work performance, productivity, and workload (RQ1), followed by the key factors that impacted the user's experience with the LSR (RQ2).

\subsection{Effect of LSR on Work Performance, Productivity, and Workload}
         % We make a clear differentiation between all three of these variables in our study. Work performance in our implementation refers to how well the participant performed in the N-back task, productivity refers to the measure of how much work the participant was able to accomplish or in other words how much faster our participants were able to reply to the messages and cognitive demand was assessed using the NASA TLX measure. Performance relates more to quality and productivity to quantity. 

         We evaluated the work performance, productivity, and workload before using LSR and after to assess how LSR impacted the collaborative work. 

                \begin{figure}[h]
                    \centering
                    \includegraphics[width=0.8\columnwidth]{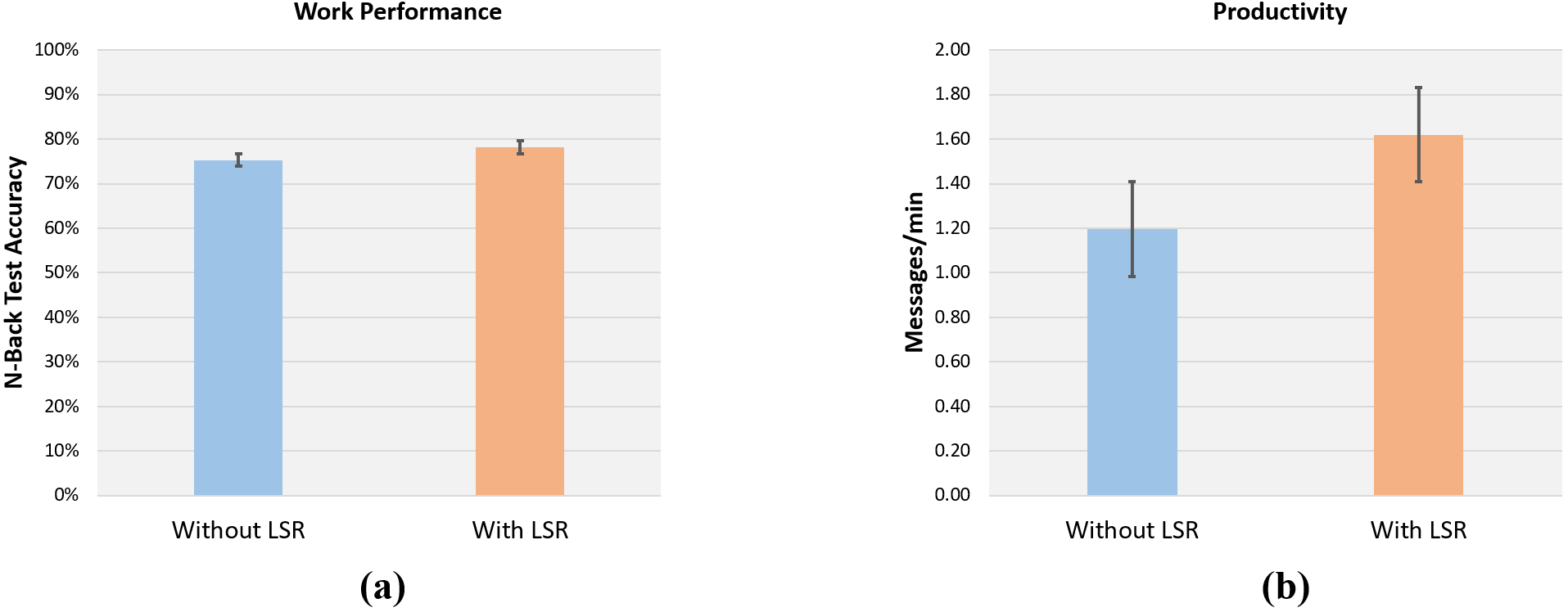}
                    \caption{Result of LSR test in collaborative work. (a) Work performance test with LSR; (b) Productivity test with LSR.}
                    \label{fig:workperformance}
                \end{figure}

\subsubsection{LSR Improves Work Performance}
    After measuring the average correctness of the N-back test as the work performance, we observed a mean score of 73.79\% and SD of 14.93\% \textbf{without LSR}, and a mean score of 79.37\% and SD of 8.96\% \textbf{with LSR}. We observed\textbf{ 5.58\%} of the difference between the two conditions. We performed the paired t-test with a t-value of \textbf{-2.56} and a p-value of \textbf{0.025} < 0.05 thus achieving a statistically significant difference. This strongly suggests that using LSR helped \textbf{increase} the work \textbf{performance}, as shown in Figure \ref{fig:workperformance} (a).
    
    % Even though the scores with AI were relatively higher compared to those without LSR, 
    % P6 felt that their performance level \textit{"would have increased bit higher if the work was not monotonous, because it was very monotonous"}. 
    P4 experienced LSR to help \textit{"improve on time management"} and even though the LSR \textit{"didn't help [them] with the task, it helped [them] get back to the task faster"}. Comparing the results of the N-Back test before and after the LSR, P8 noted even though the \textit{"work is really hard, the AI helped [them] a lot with quick responses, which save [them] a lot of time"}. 

    Based on the collected data, we conclude that the LSR may not improve work performance directly since the proposed framework does not help the participant recognize the N-back letter nor help them organize their schedules, but more likely that LSR can effectively reduce the distraction that is unrelated to work so they can be more focus on the current task, hence improve the work performance.

\subsubsection{LSR Improves Productivity}
    We calculated the average messages per minute for both conditions. These can now be considered as the actual productivity of the participants. 
    We obtained a t-value of \textbf{-7.33} and a p-value of \textbf{3.74e-06} which is statistically significant. Thus, we conclude that using LSR in these scenarios increases the overall task productivity. The result indicates an average \textbf{40.36\%} \textbf{increase} in \textbf{productivity} with LSR, as shown in Figure \ref{fig:workperformance} (b).

    % \sout{Most of our participants agreed that the LSR increased their work productivity, particularly in terms of the speed at which they could reply to their teammates and then return to the task at hand.} \hw{We missing the accurate number to conclude this}
    
    The interview suggests that the LSR increased their work productivity, particularly in terms of the speed at which they could reply to their teammates and then return to the task at hand.
    For instance, LSR greatly eased them mentally by not requiring them \textit{"to think about what to reply to the co-worker [as] the AI automatically generates all those and [they] can focus on [their] work"} [P5]. 
    Interacting with this system made them realize \textit{"how multitasking can be made easy by using AI"} [P5]. LSR greatly helped reduce cognitive demand while responding to new messages that would otherwise interfere with the user's working rhythm. As noted by P1, responding without LSR \textit{ "cost me like more work because I want to finish my task that is going on there, and I wanted to like, remember what is going on there, and when I had to like, do it in between}." In contrast, without LSR, P10 felt it was \textit{"little bit challenging to focus on ... and type at the same time"}.

        \begin{table}[h]
            \caption{Statistics for Task Load}
            \begin{tabular}{|c|c|c|}
                \toprule
                Metric                & t-Statistic& p-Value\\
                                \midrule
                Mental Demand         & 2.7102& \textbf{0.0154}\\
                Temporal Demand       & 3.6794& \textbf{0.0020}\\
                Performance& -2.5156& \textbf{0.0229}\\
                \bottomrule
            \end{tabular}
            \label{TABLE:TLXDATA}
        \end{table}

\subsubsection{LSR Reduces the Mental Demand}
\begin{figure}[h]
                \centering
                \includegraphics[width=0.8\columnwidth]{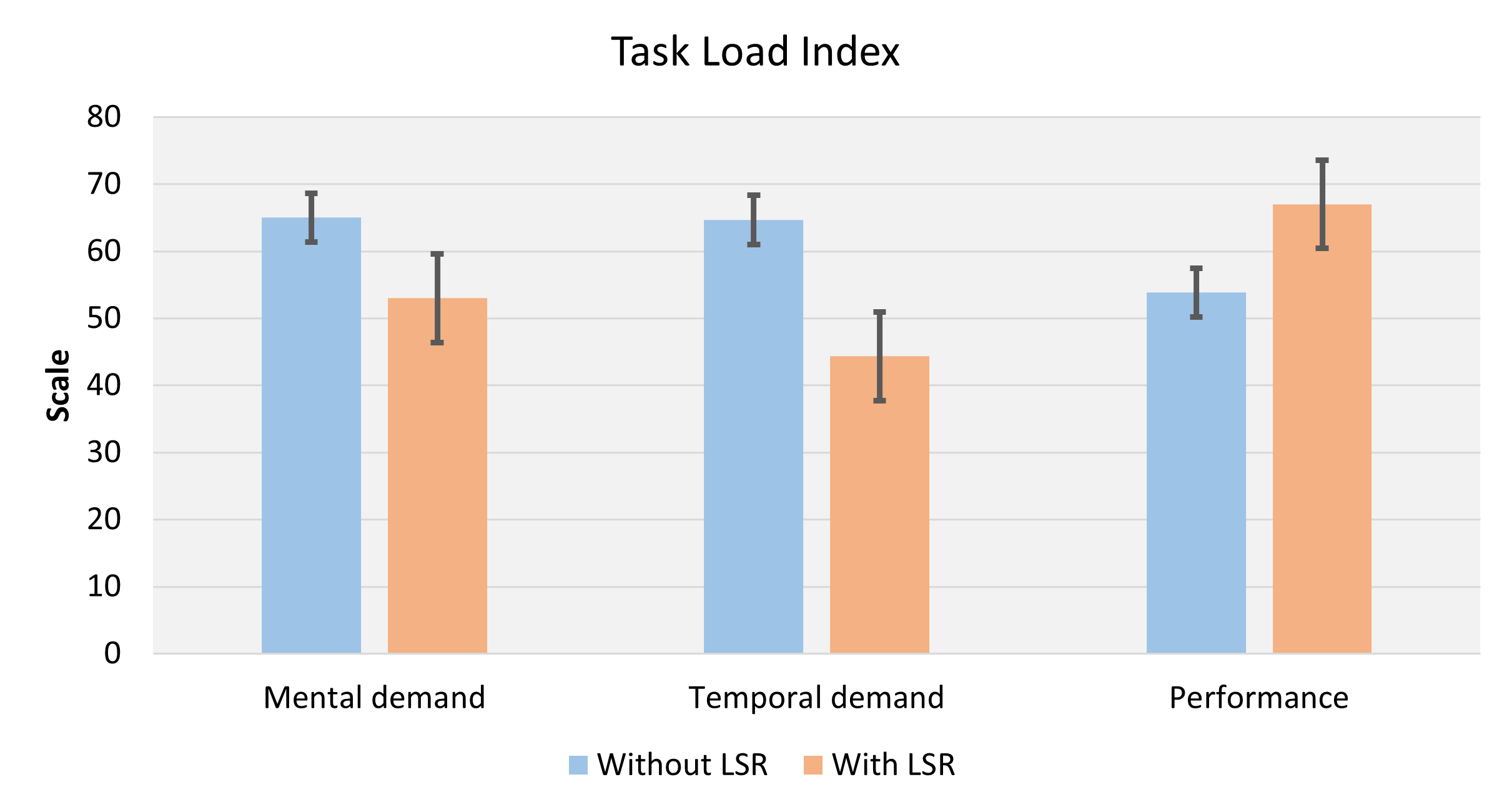}
                \caption{TLX Comparison}
                \label{fig:Grouped_TLX}
            \end{figure}

    Analysis of the NASA TLX data resulted in the following metrics displayed in Table \ref{TABLE:TLXDATA}. Results indicate that perceived \textbf{performance}, \textbf{temporal demand}, and \textbf{mental demand} are all significantly altered by the introduction of LSR. This indicates that users believe they perform the task \textbf{better} with LSR, which is consistent with our finding in the previous section, as shown in Figure \ref{fig:Grouped_TLX}.
    % Mental Demand decreased due to the lessened amount of time needed to hold in memory the letters from our N-Back test. If a user can simply click a button that allows them to return to their work quicker than needing to craft a unique response and type it themselves. The effort and Frustration metrics indicate that participants struggled to complete the N-Back task, and this is to be expected since it is a cognitively taxing task. 

The result suggests that LSR had a positive impact on the Mental Demand. The mental demand was reduced since users could quickly return to their work by clicking a button to generate responses instead of crafting them manually. 
This approach alleviated the cognitive load associated with holding information in memory, resulting in a more efficient workflow. 

    % However, the statistic result indicates Effect of Temperament does not significantly related to the assist of LSR.

 With various communication style variations from the researchers, one aspect that we expect to help reduce mental demand through the near-constant communication reply from the participant, which was 'polite' in our prompt design. When this kicked in, the users felt that LSR could answer \textit{"probably more politely than what [P3]’d react to some of the questions"}. 
 
  We embedded this inherent politeness of the LSR within our prompt that guided every output message, which the participants appreciated. P9 felt that the system provided \textit{"automated answers that you generally think in your mind. The answers were not extraordinary and in that sense."} 
Users noted they would be more likely to use these systems \textit{"when the co-workers were being rude, [because it] has the polite form of response which is good for your work environment."} This was particularly true when coworkers were \textit{"more conversational ones,"} where [P2] used it but \textit{"kind of feel bad using it because it made [P2] feel like I wasn’t paying attention to the conversation}." In contrast, other participants, such as P13, felt \textit{"more comfortable rather than typing on [their] own"}.
 P12 also felt that the AI response \textit{"was much better than what I was going to draft, so it was a good support from AI Assist in replying to those being polite at the same time while replying also."} 
    P11 voiced that the messages were not good enough in case of some formal situations, and they felt they were \textit{"not happy with the responses that were generated with Tony specifically}."

\subsection{Key Factors that Impact the User Experience}

\subsubsection{AI Performance}

\begin{figure}[h]
                    \centering
                    \includegraphics[width=1\columnwidth]{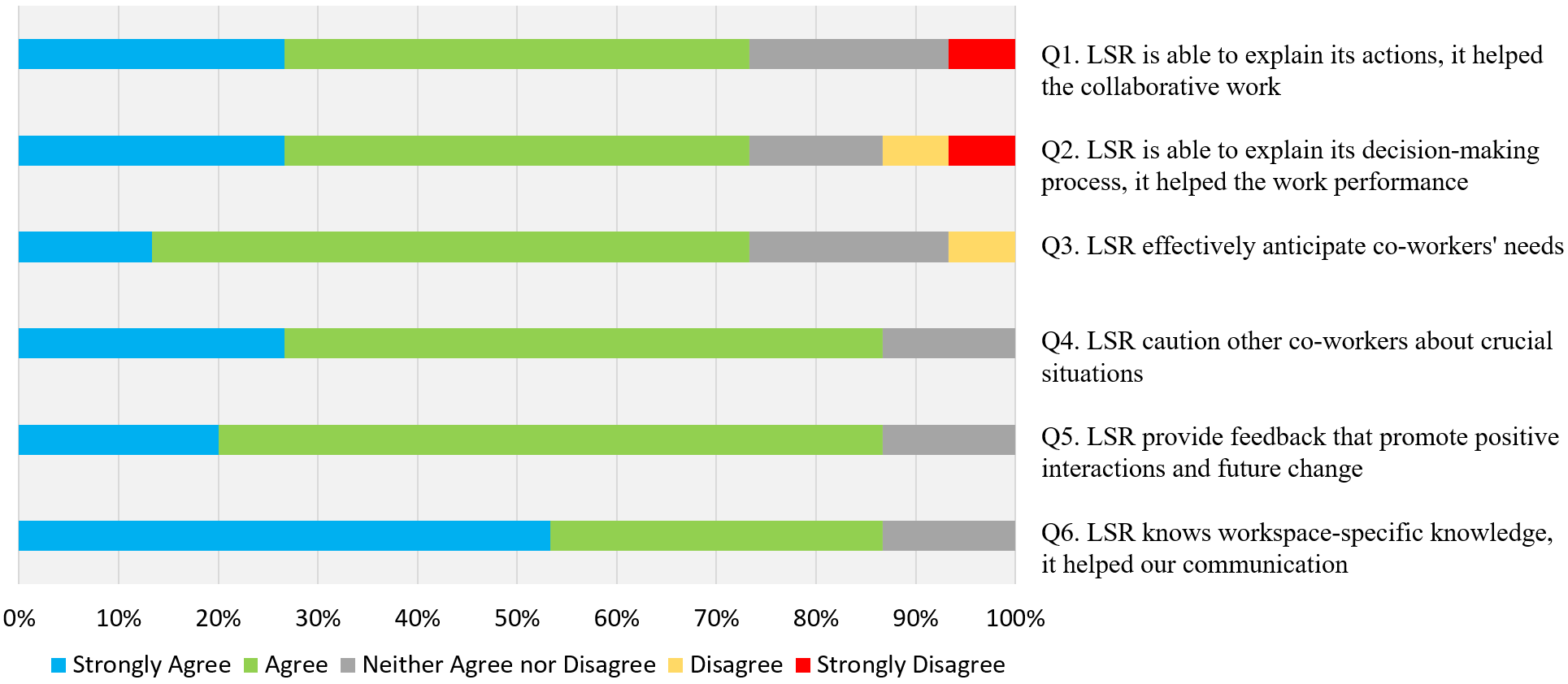}
                    \caption{Questionnaire for Measuring the General Response to LSR}
                    \label{fig:survey1}
                \end{figure}

        The survey results are presented in Figure \ref{fig:survey1} displays the distribution of responses for \textbf{Questions 1-6}, revealing that participants generally agree (>70\%) that LSR can be helpful.

    As noted by P12, \textit{"There are multiple ways to say something so you can choose from this instead of just thinking ... I was able to concentrate more on that game instead}." 
    Similarly, P11 noted, \textit{"Some of the sentence formations were good like I was not able to type that fast or type in that short time I was not able to form the sentences in the correct way, but the AI assist helped in that, and it typed it correctly or mentioned it correctly so that was the thing that I felt was unique about it or I personally felt that was great."} 
    P10 felt the same way, \textit{"There was plenty of options which was reliable to the conversation."}
    P13 highlighted \textit{"For person who ask a lot of questions it’s better to use the AI."}

     However, not all participants have high expectations toward LSR.
     For example, P10 replied, \textit{"It just like changed based on what the AI generated and I just automatically sent that message."} Our minimum implementation of the LLM model led to a communication shift due to the lack of customization.
        Some even expressed disappointment about the generated responses being misleading because \textit{"AI might have different interpretation of the word"}[P4]. As demonstrated in Figure \ref{fig:misleading_conversation}, the AI teammate replied "Glad to hear" when P4 selected the 'Good' option. In this case, our two-step process of providing a direction to a longer conversation and directly sending suffered with the AI not generalizing well to the context. This affected the participant's trust in using the system, and we observed the participant read all the outgoing generated messages thoroughly afterward, thus greatly reducing the full benefit from the system.

\begin{figure}[h]
                    \centering
                    \includegraphics[width=1\columnwidth]{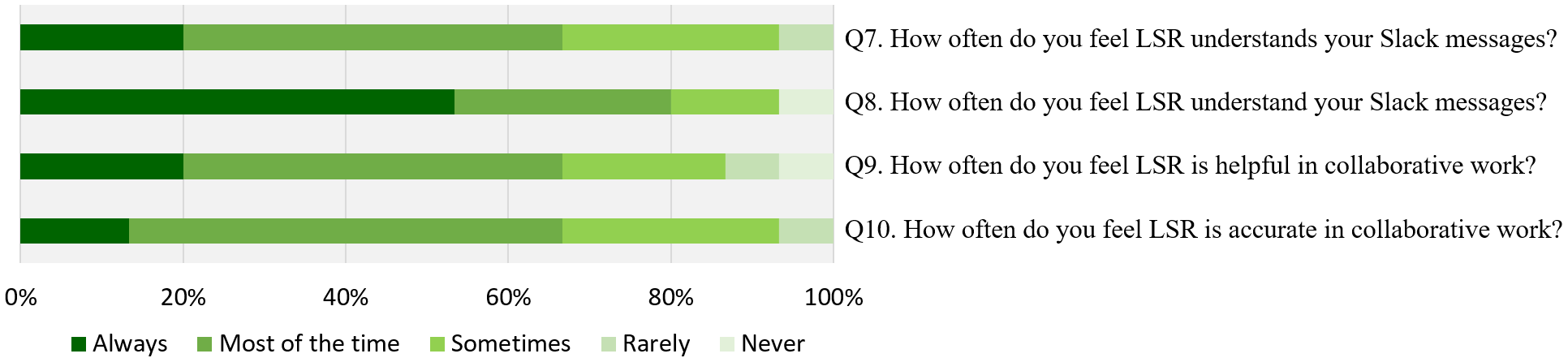}
                    \caption{Feedback of Using Experience of LSR}
                    \label{fig:survey2}
                \end{figure}

        On the other hand, Figure \ref{fig:survey2} indicates that most participants (>60\%) had a positive feeling about LSR, as they found the Slack messages accomplished by LSR are easy to understand. 

    As explained by P4, \textit{"I would have a particular response to say, but the bot would present three different alternatives which would sometimes I think that would be a better thing for me to say}."
    Regarding informal and specialized communication situations, P6 elaborated: \textit{"AI does better job in normal day life or interacting in a normal conversation, so if you’re doing a casual conversation, the AI does a good job but let say if we are doing something deep like talking to my teammates on research on proteins, the AI may not be able to address all the concerns in my chat. So I’d say AI can do good but not the best}." Similarly, the use of AI was observed to be significantly agnostic to the communication style of the other end, as explained below. 
    The use of LSR blended naturally within their conversation and developed a sort of personal bonding with the participants as explained by P6, \textit{"i know its a tool but it felt like it’s a member as well i’d say."} P6 further explained, \textit{"i was happy that AI was able to address the emotional things that I’d have replied, the AI did that and I would do exactly the same thing if I had to do it manually with my friends."}

\subsubsection{User Interface}

            With AI aiming to simplify the conversations, there exists a trade-off between granular control and independent user conversation which P2 experienced to be \textit{"slightly frustrating [because] ... instead of thinking to provide [a] response, I had to more like pick choices, kind of made me think of like speaking another language and I know what I want to say, but I have to say in other way and that kind of can be frustrating when I wanna say what I wanna say"}.
            P8 on the other hand, liked \textit{"that you could select stuffs. That works with me like agree and disagree. But it made it all nice it said something really I guess fancy. I'm pretty sure I clicked agree and it said sounds good, see you on thursday. I like that"}. 

        Issues regarding the generated messages were commonly related when using our simple two-step approach of clicking ephemeral messages to send lengthier messages. 
        For instance, P13 reported, \textit{"The answers are sometimes not so related. I got confused about the answer and just choose that"}. 
        We thus asked participants to suggest a design modification to the interface that could facilitate the message edit options after the initial selection is made to meet their needs as closely as possible. As P4 noted, \textit{"the bot would be more articulate than I would have been in some cases, but again it was lacking in some aspects and I had to stop and respond back."} 

% The interface serves as the bridge between users and the underlying AI technology (GPT-3.5 in our case), necessitating a design that’s both aesthetically pleasing and functionally effective. 
        A similar response was observed from P7, \textit{"I had problem while replying by AI-Assist because I needed to cross verify and the statements were not exactly the same that I was clicking. You know even when I click the statement is different so I’ve to cross verify what I said and there was one of them where it was Jeff or Tony, it gave a wrong day which was not intended. So this made me wait and clarify myself the time on Thursday or some other day."}

\subsubsection{Trust, Privacy, and Accessibility}
            
        The evolution of smart reply systems demands a delicate balance between user convenience and privacy to establish user trust. P14 articulately captures a common concern, remarking, \textit{"Every time you have a privacy concern regarding any AI Assist or anything related to AI."} 
        As these technologies progress, the importance of addressing inherent privacy concerns becomes even more pronounced. In evaluating real-world adoption scenarios, P8 expressed a willingness to engage with the system \textit{"at some point if [P8] trusted it enough."} Similarly, P6 voiced cautious optimism, suggesting they would \textit{"explore some more of the AI and decide after that,"} emphasizing a need to \textit{"explore a bit more before making any decision on whether [they] want AI to reply to [their] conversation or not."} Our analysis indicated a possible association between the perceived efficacy of AI and trust, which can be explored in future research. Moreover, the specific usage scenario emerged as a pivotal determinant influencing user privacy apprehensions. P4 offers an insightful perspective, noting the relative insignificance of privacy in professional contexts, asserting, \textit{"depending upon the conversation, I don't think it matters for work stuff."} This sentiment was echoed by multiple participants who expressed less privacy concern for work-related communications. Further elaborating on this, P8 added, \textit{"Not in this setting. In the work setting, no, because everything is company’s property, but in a private setting."} 
        
        Some of the major privacy concerns were regarding casual conversations and some of the workplace-specific scenarios, as presented previously. A few participants presented ideas of switching modes between AI-Assist and private mode so that smart reply functionality can be turned on or off depending on the type of conversations as mentioned by P12, \textit{"so what we can do is we can just go in a mode where ai assist could be used."} Some of our participants, \textit{"don't like buttons, [P13] prefer to talk or use voice or something like that."}

        A geographical facet was brought into the discussion by P9, who remarked, \textit{"It may be because people here in the USA are quite concerned about their private messages, so it may be an issue."} P10 voiced uncertainty about potential privacy threats, musing, \textit{"I’m not sure. It will be recorded all on the system right, the conversation? So I’m not sure about privacy."} The opacity surrounding system operations and data storage exacerbated users' privacy fears, even though AI-generated responses were based on users' prior inputs. P12 raised a valid point about the potential misuse of private information in work settings, suggesting, \textit{"if there’s a confidential project going on and we're talking about that on a chat, then retaining that information can be dangerous. We would disclose our company’s internal secrets to some third party."} Additionally, P16 emphasized the crucial role of AI service providers in ensuring stringent privacy protections, observing, \textit{"There would be much more privacy-related issues in the real-world scenario if some company wants to use this. If the provider can ensure privacy, then it could be beneficial."} Furthermore, some participants believed that privacy \textit{"could be a concern but depends on the developers of AI"}[P6]. Interestingly, some participants, like the one who noted, \textit{"while I was doing it I didn’t think of it. But when you asked me this question, I’d say yes,"} only became consciously aware of potential privacy issues upon reflection, suggesting the complex nature of users' relationship with AI-enhanced communication platforms. P13's assertion that "\textit{When I use AI, definitely I have to trust. When I use my cellphone I’ve to trust the manufacturer. There’s no way}" underscores the deep-seated trust users place in the companies that design and develop these technologies. This illuminates the imperative need for robust safeguarding mechanisms to uphold and protect user expectations and reliance on these systems when they're deployed in scale.

\subsection{Opportunities for Improvement of the LSR}

        To address the possible issues of the proposed LSR, we review the interviews with participants to explore in what instances the LSR failed to operate and why it did not meet the expectations of users in these instances. 

\subsubsection{Misinterpretation}
    An interesting case presented by P9 includes AI's inability to account for user feelings: \textit{"AI can’t answer some kinds of questions. How do you feel kind of questions, AI can’t respond to that}." A similar issue was noted by P11, \textit{"in the case of Tony, where I need to explain my feelings or I need to express my things, that AI teammate was not able to do it due to certain limitations."} We acknowledge this as a fundamental problem of human-AI interaction which is something that can only be answered with the user's response. 
    P9 further explains how the use of emojis can fail the response of LSR: \textit{"Actually in AI system if you want to ask a question and if they reply in short messages you can’t actually know if they are happy or sad or dissatisfied with you.}"

\subsubsection{Long System Response}
    Other concerns were regarding the inference time of LLMs as elicited by P8, \textit{"It was a little slow in replying even though you clicked on it."} Thus, we propose to formulate a two-step inference mechanism where a fast but less intelligent model can first help generate the ephemeral text, which the user takes a few seconds to analyze, during which the more powerful model can generate the entire sentences. This might be a useful strategy to prevent inference time to withhold the use of large-scale smart reply systems without sacrificing the user experience. 

    Another major issue with smart reply systems on large-scale deployment includes both sides using AI to reply to messages from one another. This AI-AI interaction can go over an infinite loop if proper measure is not taken, especially regarding workplace communication. In this regard, P12 explains, \textit{"if both are using AI to talk then on the same topic it can go like for a longer time until a new topic comes in because the new topic is something like a starting point for the next conversation."} P12 further suggests having a terminating mechanism to get back to work quickly instead of being trapped within the loop by generating phrases such as \textit{"I’m busy right now can we talk later, hope you don’t mind something like this. So that they can stop then and there not to elongate."}

\subsubsection{Lack of Message Customization}
        We observed many users to have their own opinion of the number of options and lengthier messages. As P10 noted, \textit{"I think its developing and may be add like more options or more detailed replies may be."} Others were fine with the default configuration, and some even desired to have \textit{"big lengthy one paragraph question and answers."} 
         As noted by P11, \textit{"AI should know about the hierarchy of those people in the company... Because like, Janine was I think a colleague then Jeff was also a colleague but that Tony was like a manager. So I think if it is having that information with AI then it would be better like those keywords or those sentences will be formed accordingly. The vocabulary used for both the cases should be different, more formal with Tony and more informal with Jannine maybe."}  
         
        After iterative analysis of the design recommendation from all of our participants, we concluded more closely with the solution proposed by P11 while also considering all of these related issues, \textit{"What I think can be done is AI assist will give you certain small tabs and if you go hover over the tab you’ll get a sentence or something like that. At least a short sentence or some keywords from the sentence maybe so that it will be good to at least find some better keywords and choose that sentence or something like that}."

\subsubsection{Multiple Message Resolving Issues}
\begin{figure}[h!]
                \centering
                \begin{subfigure}[b]{0.3\textwidth}
                    \includegraphics[width=\textwidth]{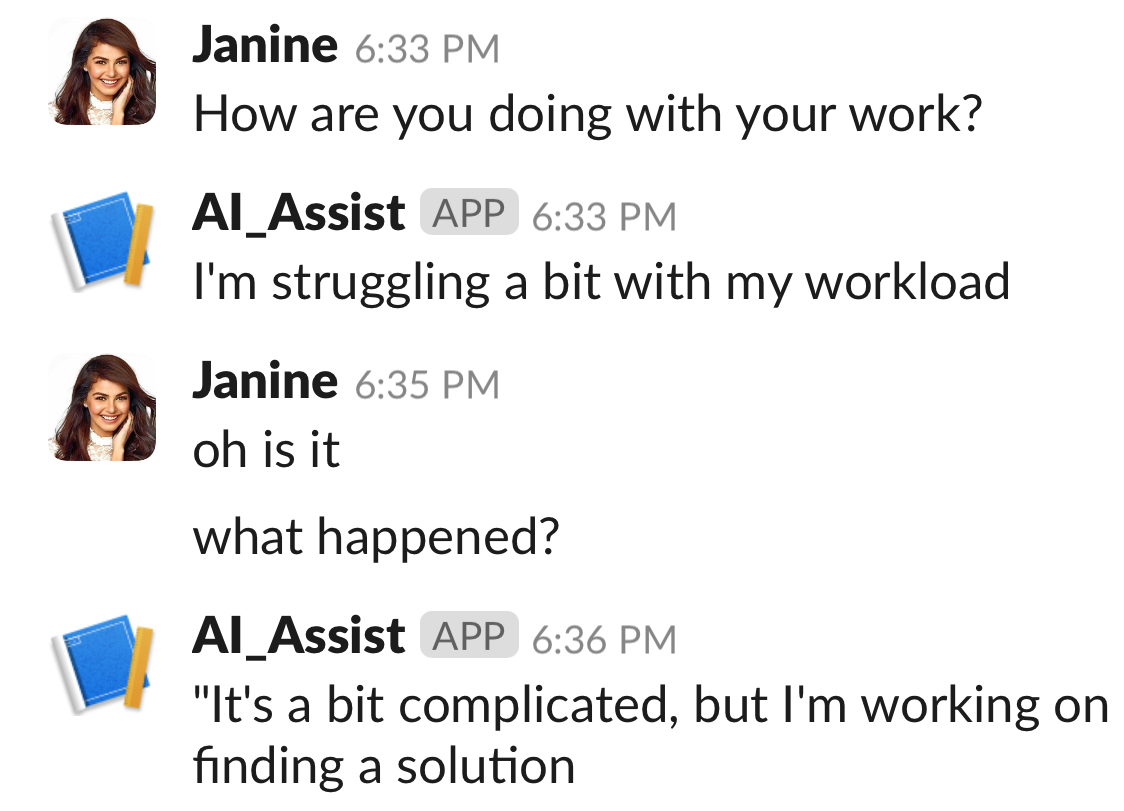}
                    \caption{Basic Working}
                    \label{}
                \end{subfigure}%
                ~
                \begin{subfigure}[b]{0.3\textwidth}
                    \includegraphics[width=\textwidth]{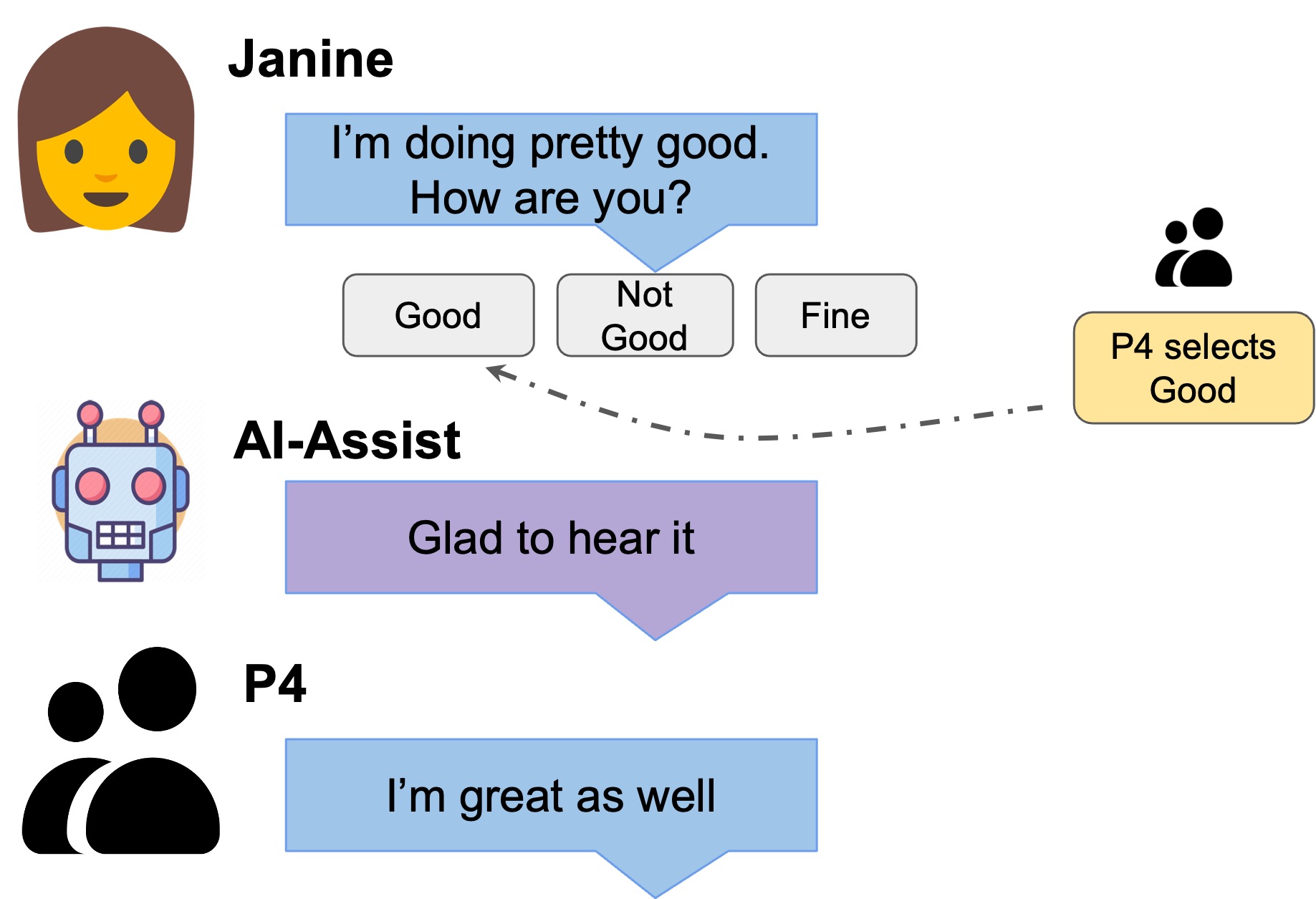}
                    \caption{Misleading Conversation}
                    \label{fig:misleading_conversation}
                \end{subfigure}%
                ~ 
                \begin{subfigure}[b]{0.3\textwidth}
                    \includegraphics[width=\textwidth]{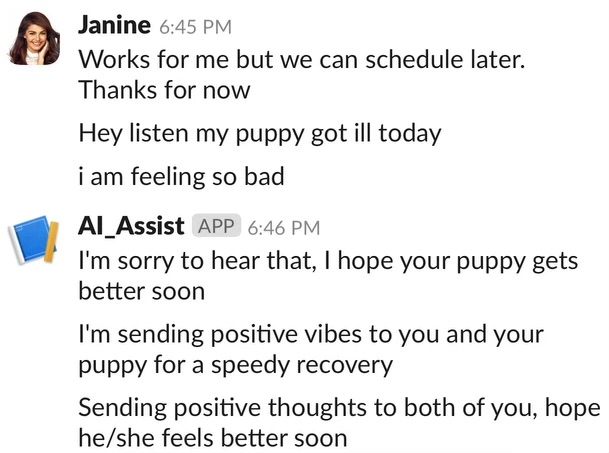}
                    \caption{AI replies to multiple messages}
                    \label{fig:multiple_messages}
                \end{subfigure}%
                ~ 
                \caption{Figure (a) shows the basic workings of how our smart reply implementation works. Figure (b) demonstrates LSR replying to the first sentence and missing out on the second question, "How are you?". Figure (c) shows complications when handling multiple asynchronous messages}
                \label{fig:three images}
            \end{figure}

         In case of multiple responses being received, we noticed a trade off of the participant being able to account for only a single message at any given time while new messages are upcoming. As noted by P12, \textit{"So I was trying to reply to this. It was related to meeting but this is the one which I replied actually ‘sending positive thoughts to both of you.' It was something that I was not intending to. This is the reason I ask for editing the response because we don't know what is exactly wrong."}
         Similar issue was observed with P7: \textit{"If someone sends me multiple messages, the AI can only reply one by one."} The LSR was found to be able to reply to only one specific message at a time, resulting in the need for multiple replies for several messages, as depicted in Figure \ref{fig:multiple_messages}. 

\subsubsection{Conversation shift}
            Another issue that we encountered was a communication shift towards some random directions of AI's generated responses leading to the user being lost within the context as was observed with P15: \textit{"when I was using the AI it automatically generates a message so I kind of sometimes lost in the context when i was talking before."} Similarly, P13 described an instance when \textit{"AI changed my answer here, I wanted to say yes it was great but AI changed it completely"} with gradual small changes summing up to something completely different.

\subsubsection{Message Inconsistency}
         Another issue might be the option sometimes not being closely representative of the response that the sender wants to formulate. P9 also suggested a similar recommendation of getting different options using some regenerate mechanism: \textit{"Like in ChatGPT if you type the same question 2 or 3 times by changing a single word you’ll get a different answer. But if we have that kind of mechanism in this program as well, then it seems the person is replying rather than AI."} 
         Since ChatGPT is inherently probabilistic, every time for the same question, we get a different response, and thus we can generate new options, which makes the addition of a single word part trivial in our case. 
         This also accounts for another concern from the same participant on having a different option than mentioned, \textit{"sometimes I need to add two or three options. Suppose three options were there and for a particular question, I was thinking of adding two options for that to answer. I mean to say ‘yes’, ‘no’ ‘maybe’, three options are there, sometimes you can say ‘yes’ or ‘maybe’. You don’t like to say no. And that option is there, then that would be better."}

\subsubsection{LSR Should Connect with Other Applications}
    % AI capabilities relate to the engine that powers these systems, responsible for delivering accurate and contextually relevant responses. We use one of the world's most powerful language model in our smart reply system. Even with this superhuman level intelligence, the participants encountered various issues regarding the system not being able to understand some of the received previous messages as well as not generlizing naturally to the context. 
    Participants also requested the LSR to access Google Calendar and other personal information to inform its responses and provide better assistance:\textit{"maybe if the AI had access to the calendar, maybe it would give replies based on whether it would be possible or not instead of me going to the calendar manually"} [P5]. 
    % P11 however suggested an interesting idea on temporally changing unidentifiable representation to tackle this issue by encoding the user information in certain form so that the user information about who and with whom remain the schedule is setup is undisclosed while also confirming who is who, \textit{"if we have some like 5 people then just give 1, 2, 3, 4 and 5 numbers to each of them, just creating the codes with respect to those persons and that information should be random and that should not be like 1 is Tony  2 is something like that. That should be like temporary information instead of having it as a permanent information. It should know that I have a meeting with same person not the other one so that's it."}

% \ab{hao please take care of this sorry}

\section{Discussion}
% As this study designed and evaluated a ChatGPT based smart reply system to support team conversation in a collaborative workplace scenario. 
% By looking at two major design aspects of interface and AI capabilities we cover the holistic overview of how these future systems can be developed so as to facilitate the user needs in a workplace environment. 
% Findings from this study contributions to the literature by filling an empirical gap as one of the first studies to investigate how the LSR impacted work performance, productivity, and workload in a collaborative workplace, with a focus on enhancing responses in formal as well as informal conversations. Based on findings from this work, in the following section we discuss the implications for design of our findings and future research directions to support better collaboration so as to increase productivity, efficiency and reduce cognitive workload while interaction with future LSR systems in collaborative work environments.

Our study explores interface design and AI capabilities to understand how future systems can better support user needs in the workplace. We discuss the implications for system design and suggest directions for future research.

\subsection{Interface Design Considerations}
% We identified several issues faced by our participants while using our bare minimum implementation. 
Our study utilized a minimalistic AI chat interface to identify user needs and potential improvements through direct interaction and feedback. Key recommendations that emerged from participant feedback are shown in the below section.

                \begin{figure}[h]
                    \centering
                    \includegraphics[width=0.8\columnwidth]{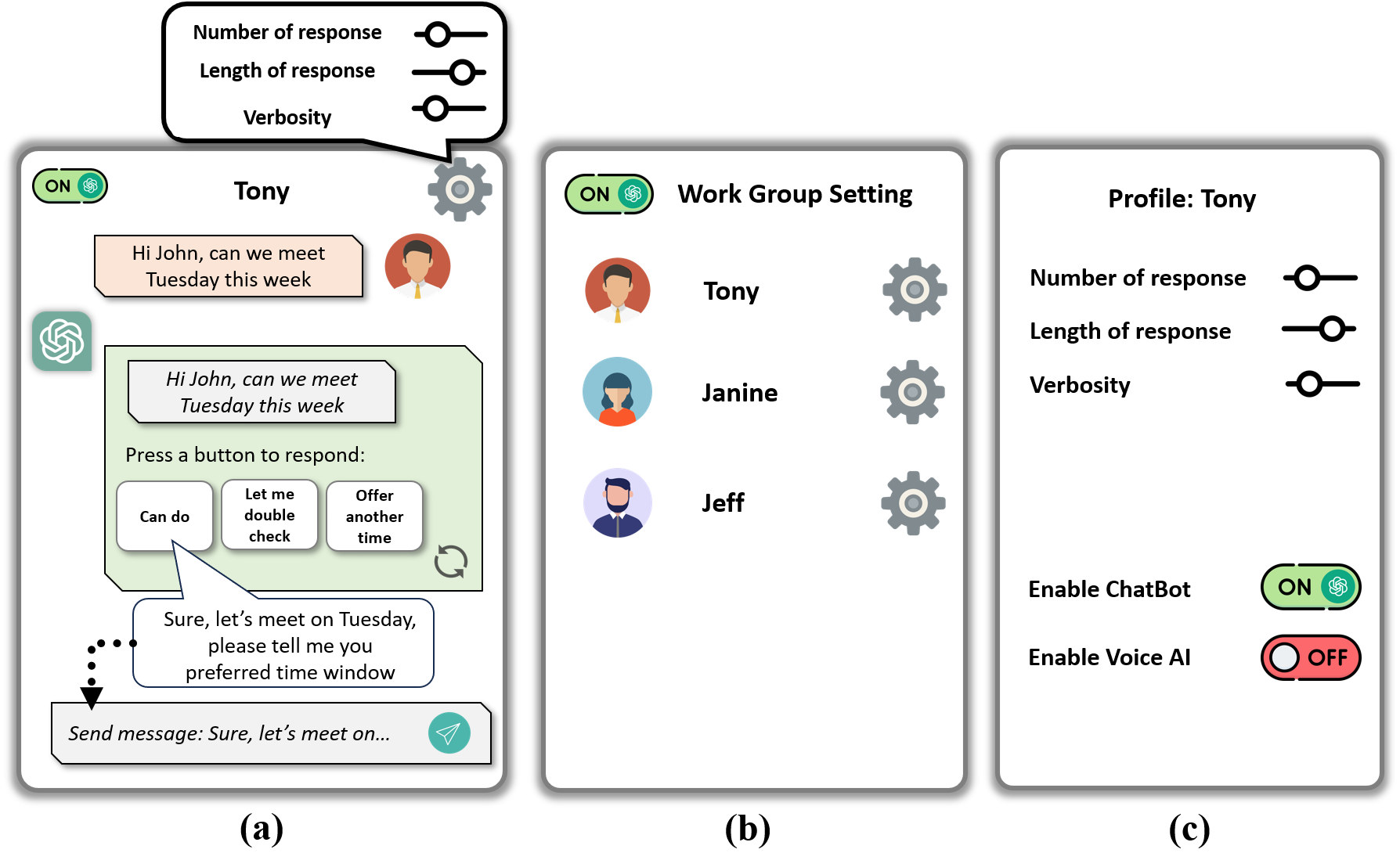}
                    \caption{An improved version of LSR based on the user feedback.}
                    \label{fig:interface1}
                \end{figure}

\subsubsection{Hover Functionality} 
Users suggested a hover-over preview of AI-generated text options to prevent misinterpretation, as illustrated in Figure \ref{fig:interface1} (a). This approach addresses clutter and ambiguity without overwhelming the interface.

\subsubsection{Regenerate and Merge Options}
For instances where AI responses were not suitable, a "regenerate" button (referenced in Figure \ref{fig:interface1} (a)) was proposed for refreshing options. Additionally, merging responses (e.g., 'yes' and 'no' into 'maybe') was recommended for more detailed communication.

\subsubsection{Customization for Workplace Communication}
Adapting AI responses based on co-worker profiles and hierarchy and personalized settings for message style and priority were identified as crucial for workplace scenarios, as shown in Figure \ref{fig:interface1} (b), (c). This customization aims to make conversations more relevant and efficient.

\subsubsection{Avoiding AI Communication Loops}
We recommend implementing a termination mechanism to prevent endless conversations between AI systems. This would limit the duration or number of messages in a conversation, requiring user input to continue or adjust the relevance of the discussion. This approach aims to keep AI interactions focused and prevent loss of context, especially after periods of disengagement.

\subsection{Trust, Privacy, and Accessibility}
Our study also highlighted several areas for improvement in AI's capabilities, focusing on enhancing user interaction and privacy:

\subsubsection{Emotional Intelligence}
Addressing AI's current limitation in understanding complex emotions and scenarios, we propose a feature that prompts users for concise inputs about their feelings. This would allow the AI to generate responses that better reflect the user's emotions and intentions.

\subsubsection{Access and Privacy}
While direct calendar integration offers convenience, it raises privacy concerns due to potential data leaks. We suggest creating a secure channel within the AI system for handling sensitive information, using anonymized and temporally encoded data to protect user privacy while maintaining functionality.

\subsubsection{Accessibility Enhancements}
Privacy concerns led to a toggle between private and AI modes (Figure \ref{fig:interface1} (a), top left). Voice-based commands were also suggested to improve accessibility, especially for users in scenarios such as driving, enhancing usability without compromising safety.

% \subsection{Limitations}
\subsection{Considerations on Experimental Design}

Our study mainly addresses the impact of Large Language Model-based Smart Reply (LSR) systems in formal collaborative workplaces, recognizing several possible limitations that provide insights for future research.

\subsubsection{Participant Diversity}
% The study's modest participant pool is a considered choice, stemming from the relatively long test time and aims to capture LSR's impact on depth over breadth.

This study focuses on the impact of LSR in the formal collaborative workplace, we recruit all participants through the university email network.
Nevertheless, our participant pool not only included people from professional workplace-specific groups, but also individuals who infrequently work in such settings (e.g., P02, P03, and P09), for more robust analysis.

Considering further expansion can help us assess LSR's effectiveness and adaptability across various scenarios beyond intensive workplace utilization. 

% Despite its limited size, statistical significance was achieved, suggesting the adequacy of the participant number for this specific research context. This approach is further reinforced by comprehensive qualitative methods, including interviews and surveys, which add further context and evidence to the quantitative findings. Moreover, the intensive nature of the experiment, requiring substantial engagement from three synchronous researchers, inherently limits the feasible sample size due to the scheduling difficulty between three researchers and participants in the limited time frames of availability. Recognizing this, the study provides insights within its scope. 

\subsubsection{Feedback from Recipients}
Although this research primarily focuses on the sender's perspective, it is also essential to develop complementary works to assess the recipient's experience as well. 
A critical consideration is the impact of LSR on the quality of interaction. While facilitating efficient communication, LSR might prioritize task completion over meaningful engagement, potentially affecting the authenticity and personalization in professional relationships. 
This aspect becomes particularly relevant when long-standing work relationships require a deeper understanding and a personal touch. 

% For instance, some applications now offer auto-generated responses (e.g., Gmail), aiming to boost communication. However, when recipients receive these semi-automated responses, the interaction can feel impersonal and transactional. 
% Although the proposed LSR is built to eliminate this insecurity, this perception can subtly erode the sense of personal connection and appreciation that would be conveyed through a customized message.

\section{Conclusion}

Our research investigated the use of an LLM-based smart reply system to enhance conversation in the workplace. This system aims to provide context-aware, personalized responses. 
In a simulated work environment, findings revealed that LSR significantly reduces workload and improves task performance and productivity, as indicated by NASA TLX scores and N-back task results. 
Along with these, we contribute to two major sections: User Interaction and Design considerations. 
We dive deep into the user experience, trust, privacy, and accessibility, and also propose detailed design considerations based on our findings while focusing specifically on the interface (the front end) and the AI capabilities(the back end). 
These discussions reveal the possible improvements of LLMs as well as their potential issue, suggesting a direction for future trustworthy conversational agents in professional contexts.

\bibliography{Bibliography.bib}

\begin{thebibliography}{10}

\bibitem{achiam2023gpt}
J.~Achiam, S.~Adler, S.~Agarwal, L.~Ahmad, I.~Akkaya, F.~L. Aleman, D.~Almeida,
  J.~Altenschmidt, S.~Altman, S.~Anadkat, {\em et~al.}, ``Gpt-4 technical
  report,'' {\em arXiv preprint arXiv:2303.08774}, 2023.

\bibitem{katz2024gpt}
D.~M. Katz, M.~J. Bommarito, S.~Gao, and P.~Arredondo, ``Gpt-4 passes the bar
  exam,'' {\em Philosophical Transactions of the Royal Society A}, vol.~382,
  no.~2270, p.~20230254, 2024.

\bibitem{team2023gemini}
G.~Team, R.~Anil, S.~Borgeaud, Y.~Wu, J.-B. Alayrac, J.~Yu, R.~Soricut,
  J.~Schalkwyk, A.~M. Dai, A.~Hauth, {\em et~al.}, ``Gemini: a family of highly
  capable multimodal models,'' {\em arXiv preprint arXiv:2312.11805}, 2023.

\bibitem{touvron2023llama}
H.~Touvron, T.~Lavril, G.~Izacard, X.~Martinet, M.-A. Lachaux, T.~Lacroix,
  B.~Rozi{\`e}re, N.~Goyal, E.~Hambro, F.~Azhar, {\em et~al.}, ``Llama: Open
  and efficient foundation language models,'' {\em arXiv preprint
  arXiv:2302.13971}, 2023.

\bibitem{wu2023autogen}
Q.~Wu, G.~Bansal, J.~Zhang, Y.~Wu, S.~Zhang, E.~Zhu, B.~Li, L.~Jiang, X.~Zhang,
  and C.~Wang, ``Autogen: Enabling next-gen llm applications via multi-agent
  conversation framework,'' {\em arXiv preprint arXiv:2308.08155}, 2023.

\bibitem{cai2022context}
S.~Cai, S.~Venugopalan, K.~Tomanek, A.~Narayanan, M.~R. Morris, and M.~P.
  Brenner, ``Context-aware abbreviation expansion using large language
  models,'' {\em arXiv preprint arXiv:2205.03767}, 2022.

\bibitem{goodman2022lampost}
S.~M. Goodman, E.~Buehler, P.~Clary, A.~Coenen, A.~Donsbach, T.~N. Horne,
  M.~Lahav, R.~MacDonald, R.~B. Michaels, A.~Narayanan, {\em et~al.},
  ``Lampost: Design and evaluation of an ai-assisted email writing prototype
  for adults with dyslexia,'' in {\em Proceedings of the 24th International ACM
  SIGACCESS Conference on Computers and Accessibility}, pp.~1--18, 2022.

\bibitem{yuan2022wordcraft}
A.~Yuan, A.~Coenen, E.~Reif, and D.~Ippolito, ``Wordcraft: story writing with
  large language models,'' in {\em 27th International Conference on Intelligent
  User Interfaces}, pp.~841--852, 2022.

\bibitem{kamel1998applying}
N.~N. Kamel and R.~M. Davison, ``Applying cscw technology to overcome
  traditional barriers in group interactions,'' {\em Information \&
  Management}, vol.~34, no.~4, pp.~209--219, 1998.

\bibitem{carasik1988case}
R.~P. Carasik and C.~E. Grantham, ``A case study of cscw in a dispersed
  organization,'' in {\em Proceedings of the SIGCHI conference on Human factors
  in computing systems}, pp.~61--66, 1988.

\bibitem{convertino2007supporting}
G.~Convertino, U.~Farooq, M.~B. Rosson, J.~M. Carroll, and B.~J. Meyer,
  ``Supporting intergenerational groups in computer-supported cooperative work
  (cscw),'' {\em Behaviour \& Information Technology}, vol.~26, no.~4,
  pp.~275--285, 2007.

\bibitem{yadav2019feedpal}
D.~Yadav, P.~Malik, K.~Dabas, and P.~Singh, ``Feedpal: Understanding
  opportunities for chatbots in breastfeeding education of women in india,''
  {\em Proceedings of the ACM on Human-Computer Interaction}, vol.~3, no.~CSCW,
  pp.~1--30, 2019.

\bibitem{srba2015askalot}
I.~Srba and M.~Bielikova, ``Askalot: community question answering as a means
  for knowledge sharing in an educational organization,'' in {\em Proceedings
  of the 18th ACM conference companion on computer supported cooperative work
  \& social computing}, pp.~179--182, 2015.

\bibitem{enembreck2002personal}
F.~Enembreck and J.-P. Barthes, ``Personal assistant to improve cscw,'' in {\em
  The 7th International Conference on Computer Supported Cooperative Work in
  Design}, pp.~329--335, IEEE, 2002.

\bibitem{hancock2020ai}
J.~T. Hancock, M.~Naaman, and K.~Levy, ``Ai-mediated communication: Definition,
  research agenda, and ethical considerations,'' {\em Journal of
  Computer-Mediated Communication}, vol.~25, no.~1, pp.~89--100, 2020.

\bibitem{zhang2023impact}
Z.~Zhang, B.~Li, and L.~Liu, ``The impact of ai-based conversational agent on
  the firms’ operational performance: Empirical evidence from a call
  center,'' {\em Applied Artificial Intelligence}, vol.~37, no.~1, p.~2157592,
  2023.

\bibitem{lewis2016appreciative}
S.~Lewis, J.~Passmore, and S.~Cantore, {\em Appreciative inquiry for change
  management: Using AI to facilitate organizational development}.
\newblock Kogan Page Publishers, 2016.

\bibitem{dwivedi2023so}
Y.~K. Dwivedi, N.~Kshetri, L.~Hughes, E.~L. Slade, A.~Jeyaraj, A.~K. Kar, A.~M.
  Baabdullah, A.~Koohang, V.~Raghavan, M.~Ahuja, {\em et~al.}, ``“so what if
  chatgpt wrote it?” multidisciplinary perspectives on opportunities,
  challenges and implications of generative conversational ai for research,
  practice and policy,'' {\em International Journal of Information Management},
  vol.~71, p.~102642, 2023.

\bibitem{teixeira2021interplay}
M.~S. Teixeira, V.~Maran, and M.~Dragoni, ``The interplay of a conversational
  ontology and ai planning for health dialogue management,'' in {\em
  Proceedings of the 36th annual ACM symposium on applied computing},
  pp.~611--619, 2021.

\bibitem{kulkarni2019conversational}
P.~Kulkarni, A.~Mahabaleshwarkar, M.~Kulkarni, N.~Sirsikar, and K.~Gadgil,
  ``Conversational ai: An overview of methodologies, applications \& future
  scope,'' in {\em 2019 5th International Conference On Computing,
  Communication, Control And Automation (ICCUBEA)}, pp.~1--7, IEEE, 2019.

\bibitem{cifci2023ai}
D.~Cifci, G.~P. Veldhuizen, S.~Foersch, and J.~N. Kather, ``Ai in computational
  pathology of cancer: Improving diagnostic workflows and clinical outcomes?,''
  {\em Annual Review of Cancer Biology}, vol.~7, pp.~57--71, 2023.

\bibitem{li2022using}
R.~C. Li, M.~Smith, J.~Lu, A.~Avati, S.~Wang, W.~G. Teuteberg, K.~Shum,
  G.~Hong, B.~Seevaratnam, J.~Westphal, {\em et~al.}, ``Using ai to empower
  collaborative team workflows: two implementations for advance care planning
  and care escalation,'' {\em NEJM Catalyst Innovations in Care Delivery},
  vol.~3, no.~4, pp.~CAT--21, 2022.

\bibitem{bidot2011using}
J.~Bidot, C.~Goumopoulos, and I.~Calemis, ``Using ai planning and late binding
  for managing service workflows in intelligent environments,'' in {\em 2011
  IEEE International Conference on Pervasive Computing and Communications
  (PerCom)}, pp.~156--163, IEEE, 2011.

\bibitem{bastola2024driving}
A.~Bastola, J.~Brinkley, H.~Wang, and A.~Razi, ``Driving towards inclusion:
  Revisiting in-vehicle interaction in autonomous vehicles,'' {\em arXiv
  preprint arXiv:2401.14571}, 2024.

\bibitem{jiang2022beyond}
N.~Jiang, X.~Liu, H.~Liu, E.~T.~K. Lim, C.-W. Tan, and J.~Gu, ``Beyond
  ai-powered context-aware services: the role of human--ai collaboration,''
  {\em Industrial Management \& Data Systems}, 2022.

\bibitem{santos2016toward}
O.~C. Santos, M.~Saneiro, J.~G. Boticario, and M.~C. Rodriguez-Sanchez,
  ``Toward interactive context-aware affective educational recommendations in
  computer-assisted language learning,'' {\em New Review of Hypermedia and
  Multimedia}, vol.~22, no.~1-2, pp.~27--57, 2016.

\bibitem{zarate2008collaborative}
P.~Zarat{\'e}, J.~P. Belaud, and G.~Camilleri, ``Collaborative decision making:
  Perspectives and challenges,'' 2008.

\bibitem{schaefer2019integrating}
K.~E. Schaefer, J.~Oh, D.~Aksaray, and D.~Barber, ``Integrating context into
  artificial intelligence: research from the robotics collaborative technology
  alliance,'' {\em Ai Magazine}, vol.~40, no.~3, pp.~28--40, 2019.

\bibitem{kunnathuvalappil2018artificial}
N.~Kunnathuvalappil~Hariharan, ``Artificial intelligence and human
  collaboration in financial planning,'' 2018.

\bibitem{bastola2023multi}
A.~Bastola, M.~Atik~Enam, A.~Bastola, A.~Gluck, and J.~Brinkley,
  ``Multi-functional glasses for the blind and visually impaired: Design and
  development,'' in {\em Proceedings of the Human Factors and Ergonomics
  Society Annual Meeting}, vol.~67, pp.~995--1001, SAGE Publications Sage CA:
  Los Angeles, CA, 2023.

\bibitem{chen2022two}
X.~Chen, D.~Zou, H.~Xie, G.~Cheng, and C.~Liu, ``Two decades of artificial
  intelligence in education,'' {\em Educational Technology \& Society},
  vol.~25, no.~1, pp.~28--47, 2022.

\bibitem{wilson2018collaborative}
H.~J. Wilson and P.~R. Daugherty, ``Collaborative intelligence: Humans and ai
  are joining forces,'' {\em Harvard Business Review}, vol.~96, no.~4,
  pp.~114--123, 2018.

\bibitem{tien2017internet}
J.~M. Tien, ``Internet of things, real-time decision making, and artificial
  intelligence,'' {\em Annals of Data Science}, vol.~4, pp.~149--178, 2017.

\bibitem{epstein2015wanted}
S.~L. Epstein, ``Wanted: collaborative intelligence,'' {\em Artificial
  Intelligence}, vol.~221, pp.~36--45, 2015.

\bibitem{kung2023performance}
T.~H. Kung, M.~Cheatham, A.~Medenilla, C.~Sillos, L.~De~Leon, C.~Elepa{\~n}o,
  M.~Madriaga, R.~Aggabao, G.~Diaz-Candido, J.~Maningo, {\em et~al.},
  ``Performance of chatgpt on usmle: Potential for ai-assisted medical
  education using large language models,'' {\em PLoS digital health}, vol.~2,
  no.~2, p.~e0000198, 2023.

\bibitem{regan2024can}
C.~Regan, N.~Iwahashi, S.~Tanaka, and M.~Oka, ``Can generative agents predict
  emotion?,'' {\em arXiv preprint arXiv:2402.04232}, 2024.

\bibitem{Ref:Humans_can't_tell}
E.~Clark, T.~August, S.~Serrano, N.~Haduong, S.~Gururangan, and N.~A. Smith,
  ``All that{'}s {`}human{'} is not gold: Evaluating human evaluation of
  generated text,'' in {\em Proceedings of the 59th Annual Meeting of the
  Association for Computational Linguistics and the 11th International Joint
  Conference on Natural Language Processing (Volume 1: Long Papers)}, (Online),
  pp.~7282--7296, Association for Computational Linguistics, Aug. 2021.

\bibitem{Ref:Smart_Reply}
A.~Kannan, K.~Kurach, S.~Ravi, T.~Kaufmann, A.~Tomkins, B.~Miklos, G.~Corrado,
  L.~Lukacs, M.~Ganea, P.~Young, and V.~Ramavajjala, ``Smart reply: Automated
  response suggestion for email,'' 2016.

\bibitem{Ref:Smart_compose}
M.~X. Chen, B.~N. Lee, G.~Bansal, Y.~Cao, S.~Zhang, J.~Lu, J.~Tsay, Y.~Wang,
  A.~M. Dai, Z.~Chen, T.~Sohn, and Y.~Wu, ``Gmail smart compose: Real-time
  assisted writing,'' in {\em Proceedings of the 25th ACM SIGKDD International
  Conference on Knowledge Discovery; Data Mining}, KDD '19, (New York, NY,
  USA), p.~2287–2295, Association for Computing Machinery, 2019.

\bibitem{kannan2016smart}
A.~Kannan, K.~Kurach, S.~Ravi, T.~Kaufmann, A.~Tomkins, B.~Miklos, G.~Corrado,
  L.~Lukacs, M.~Ganea, P.~Young, {\em et~al.}, ``Smart reply: Automated
  response suggestion for email,'' in {\em Proceedings of the 22nd ACM SIGKDD
  international conference on knowledge discovery and data mining},
  pp.~955--964, 2016.

\bibitem{devlin2018bert}
J.~Devlin, M.-W. Chang, K.~Lee, and K.~Toutanova, ``Bert: Pre-training of deep
  bidirectional transformers for language understanding,'' {\em arXiv preprint
  arXiv:1810.04805}, 2018.

\bibitem{brown2020language}
T.~Brown, B.~Mann, N.~Ryder, M.~Subbiah, J.~D. Kaplan, P.~Dhariwal,
  A.~Neelakantan, P.~Shyam, G.~Sastry, A.~Askell, {\em et~al.}, ``Language
  models are few-shot learners,'' {\em Advances in neural information
  processing systems}, vol.~33, pp.~1877--1901, 2020.

\bibitem{bubeck2023sparks}
S.~Bubeck, V.~Chandrasekaran, R.~Eldan, J.~Gehrke, E.~Horvitz, E.~Kamar,
  P.~Lee, Y.~T. Lee, Y.~Li, S.~Lundberg, {\em et~al.}, ``Sparks of artificial
  general intelligence: Early experiments with gpt-4,'' {\em arXiv preprint
  arXiv:2303.12712}, 2023.

\bibitem{henderson2017efficient}
M.~Henderson, R.~Al-Rfou, B.~Strope, Y.-H. Sung, L.~Luk{\'a}cs, R.~Guo,
  S.~Kumar, B.~Miklos, and R.~Kurzweil, ``Efficient natural language response
  suggestion for smart reply,'' {\em arXiv preprint arXiv:1705.00652}, 2017.

\bibitem{moosa2016smart}
J.~M. Moosa, ``How smart is the “smart reply” feature of google allo: From
  users’ viewpoint,'' 2016.

\bibitem{xiao2023powering}
Z.~Xiao, Q.~V. Liao, M.~Zhou, T.~Grandison, and Y.~Li, ``Powering an ai chatbot
  with expert sourcing to support credible health information access,'' in {\em
  Proceedings of the 28th International Conference on Intelligent User
  Interfaces}, pp.~2--18, 2023.

\bibitem{mieczkowski2021ai}
H.~Mieczkowski, J.~T. Hancock, M.~Naaman, M.~Jung, and J.~Hohenstein,
  ``Ai-mediated communication: Language use and interpersonal effects in a
  referential communication task,'' {\em Proceedings of the ACM on
  Human-Computer Interaction}, vol.~5, no.~CSCW1, pp.~1--14, 2021.

\bibitem{hohenstein2023artificial}
J.~Hohenstein, R.~F. Kizilcec, D.~DiFranzo, Z.~Aghajari, H.~Mieczkowski,
  K.~Levy, M.~Naaman, J.~Hancock, and M.~F. Jung, ``Artificial intelligence in
  communication impacts language and social relationships,'' {\em Scientific
  Reports}, vol.~13, no.~1, p.~5487, 2023.

\bibitem{kim2023propile}
S.~Kim, S.~Yun, H.~Lee, M.~Gubri, S.~Yoon, and S.~J. Oh, ``Propile: Probing
  privacy leakage in large language models,'' {\em arXiv preprint
  arXiv:2307.01881}, 2023.

\bibitem{wamba2020influence}
S.-L. Wamba-Taguimdje, S.~Fosso~Wamba, J.~R. Kala~Kamdjoug, and C.~E.
  Tchatchouang~Wanko, ``Influence of artificial intelligence (ai) on firm
  performance: the business value of ai-based transformation projects,'' {\em
  Business Process Management Journal}, vol.~26, no.~7, pp.~1893--1924, 2020.

\bibitem{shim2002past}
J.~P. Shim, M.~Warkentin, J.~F. Courtney, D.~J. Power, R.~Sharda, and
  C.~Carlsson, ``Past, present, and future of decision support technology,''
  {\em Decision support systems}, vol.~33, no.~2, pp.~111--126, 2002.

\bibitem{wenker2023wrote}
K.~Wenker, ``Who wrote this? how smart replies impact language and agency in
  the workplace,'' {\em Telematics and Informatics Reports}, vol.~10,
  p.~100062, 2023.

\bibitem{eloundou2023gpts}
T.~Eloundou, S.~Manning, P.~Mishkin, and D.~Rock, ``Gpts are gpts: An early
  look at the labor market impact potential of large language models,'' {\em
  arXiv preprint arXiv:2303.10130}, 2023.

\bibitem{manyika2017future}
J.~Manyika, M.~Chui, M.~Miremadi, J.~Bughin, K.~George, P.~Willmott, and
  M.~Dewhurst, ``A future that works: Ai, automation, employment, and
  productivity,'' {\em McKinsey Global Institute Research, Tech. Rep}, vol.~60,
  pp.~1--135, 2017.

\bibitem{urquhart2022working}
L.~Urquhart, A.~Laffer, and D.~Miranda, ``Working with affective computing:
  Exploring uk public perceptions of ai enabled workplace surveillance,'' {\em
  arXiv preprint arXiv:2205.08264}, 2022.

\bibitem{kasneci2023chatgpt}
E.~Kasneci, K.~Se{\ss}ler, S.~K{\"u}chemann, M.~Bannert, D.~Dementieva,
  F.~Fischer, U.~Gasser, G.~Groh, S.~G{\"u}nnemann, E.~H{\"u}llermeier, {\em
  et~al.}, ``Chatgpt for good? on opportunities and challenges of large
  language models for education,'' {\em Learning and individual differences},
  vol.~103, p.~102274, 2023.

\bibitem{srivastava2022beyond}
A.~Srivastava, A.~Rastogi, A.~Rao, A.~A.~M. Shoeb, A.~Abid, A.~Fisch, A.~R.
  Brown, A.~Santoro, A.~Gupta, A.~Garriga-Alonso, {\em et~al.}, ``Beyond the
  imitation game: Quantifying and extrapolating the capabilities of language
  models,'' {\em arXiv preprint arXiv:2206.04615}, 2022.

\bibitem{nawaz2019artificial}
N.~Nawaz and A.~M. Gomes, ``Artificial intelligence chatbots are new
  recruiters,'' {\em IJACSA) International Journal of Advanced Computer Science
  and Applications}, vol.~10, no.~9, 2019.

\bibitem{krishnan2022impact}
C.~Krishnan, A.~Gupta, A.~Gupta, and G.~Singh, ``Impact of artificial
  intelligence-based chatbots on customer engagement and business growth,'' in
  {\em Deep Learning for Social Media Data Analytics}, pp.~195--210, Springer,
  2022.

\bibitem{getchell2022artificial}
K.~M. Getchell, S.~Carradini, P.~W. Cardon, C.~Fleischmann, H.~Ma, J.~Aritz,
  and J.~Stapp, ``Artificial intelligence in business communication: the
  changing landscape of research and teaching,'' {\em Business and Professional
  Communication Quarterly}, vol.~85, no.~1, pp.~7--33, 2022.

\bibitem{pereira2023systematic}
V.~Pereira, E.~Hadjielias, M.~Christofi, and D.~Vrontis, ``A systematic
  literature review on the impact of artificial intelligence on workplace
  outcomes: A multi-process perspective,'' {\em Human Resource Management
  Review}, vol.~33, no.~1, p.~100857, 2023.

\bibitem{owen2005n}
A.~M. Owen, K.~M. McMillan, A.~R. Laird, and E.~Bullmore, ``N-back working
  memory paradigm: A meta-analysis of normative functional neuroimaging
  studies,'' {\em Human brain mapping}, vol.~25, no.~1, pp.~46--59, 2005.

\bibitem{hart2006nasa}
S.~G. Hart, ``Nasa-task load index (nasa-tlx); 20 years later,'' in {\em
  Proceedings of the human factors and ergonomics society annual meeting},
  vol.~50, pp.~904--908, Sage publications Sage CA: Los Angeles, CA, 2006.

\bibitem{Ref:An_ideal_human}
R.~Zhang, N.~J. McNeese, G.~Freeman, and G.~Musick, ``" an ideal human"
  expectations of ai teammates in human-ai teaming,'' {\em Proceedings of the
  ACM on Human-Computer Interaction}, vol.~4, no.~CSCW3, pp.~1--25, 2021.

\bibitem{braun2006using}
V.~Braun and V.~Clarke, ``Using thematic analysis in psychology,'' {\em
  Qualitative research in psychology}, vol.~3, no.~2, pp.~77--101, 2006.

\end{thebibliography}
\bibliographystyle{ieeetr}

% \begin{table}
%   \caption{Frequency of Special Characters}
%   \label{tab:freq}
%   \begin{tabular}{ccl}
%     \toprule
%     Non-English or Math&Frequency&Comments\\
%     \midrule
%     \O & 1 in 1,000& For Swedish names\\
%     $\pi$ & 1 in 5& Common in math\\
%     \$ & 4 in 5 & Used in business\\
%     $\Psi^2_1$ & 1 in 40,000& Unexplained usage\\
%   \bottomrule
% \end{tabular}
% \end{table}

% \begin{table*}
%   \caption{Some Typical Commands}
%   \label{tab:commands}
%   \begin{tabular}{ccl}
%     \toprule
%     Command &A Number & Comments\\
%     \midrule
%     \texttt{{\char'134}author} & 100& Author \\
%     \texttt{{\char'134}table}& 300 & For tables\\
%     \texttt{{\char'134}table*}& 400& For wider tables\\
%     \bottomrule
%   \end{tabular}
% \end{table*}

% \begin{figure}[h]
%   \centering
%   \includegraphics[width=\linewidth]{sample-franklin}
%   \caption{1907 Franklin Model D roadster. Photograph by Harris \&
%     Ewing, Inc. [Public domain], via Wikimedia
%     Commons. (\url{https://goo.gl/VLCRBB}).}
% \end{figure}

\end{document}